\documentstyle[preprint,eqsecnum,aps]{revtex}

\tightenlines

\begin{document}

\draft
\preprint{}
\title{Theory of exciton-exciton correlation in nonlinear optical response}

\author{Th. \"Ostreich and K. Sch\"onhammer}   
\address{Institut f\"ur Theoretische Physik,
Universit\"at G\"ottingen, Bunsenstra\ss{}e 9, D-37073 G\"ottingen,
Federal Republic of Germany}

\author{L. J. Sham}
\address{Department of Physics, University of California 
San Diego, La Jolla, California 92093-0319}

\date{\today}

\maketitle  

\begin{abstract}
We present a systematic theory of Coulomb interaction effects
in the nonlinear optical processes in semiconductors 
using a perturbation series in the exciting laser field. 
The third-order dynamical response consists of phase-space filling correction,
mean-field exciton-exciton interaction, and two-exciton correlation effects
expressed as a force-force correlation function.
The theory provides a unified description of effects of bound and unbound
biexcitons, including memory-effects beyond the Markovian approximation. In the
degenerate four-wave-mixing experiments, correlation effects are shown leading
to polarization mixing, ringing, etc.  The strong interaction, nonperturbative
theory of the correlation function is numerically evaluated for a
one-dimensional model. 
Approximations for the correlation function are presented.
\end{abstract}
\vskip 0.1 truein

\narrowtext



%
\section{Introduction}
\label{section1}
%


Transient four-wave-mixing (FWM) experiments have proven to
be a powerful tool in probing and understanding
optical coherence in semiconductors \cite{leo,exho,sr2,weiss,kim}.
Subpicosecond spectroscopy yields information
on the very early stages of time development of the carrier dynamics and
many-particle correlations.


The essential physical picture behind the dynamical evolution
of optically excited electrons and holes can be understood
in simple terms \cite{leo,weiss,kim,2,ste,chem2}. 
First, the ultrafast dynamics of the exciting laser field with frequency near
the fundamental band gap of a semiconductor creates coherent electron-hole (eh)
pairs.  Secondly, the motion of the carriers, dominated  by the Coulomb
interaction among them, leads to an ultrafast electric polarization
as a source of light which can be observed. 
The scattering of carriers by other carriers, by phonons, and by defects
leads to polarization decay and loss of optical coherence \cite{goebel}.


The density-matrix equations of motion which described the dynamics of the
microscopic polarization and particle distribution functions were established
by a number of groups \cite{schmitt,koch88,schaf,14,stahl}. 
Within the mean-field approximation, results for the ultrafast dynamics of the
electron-hole pairs agreed well with the extant experimental findings, 
for example, for the dynamical Stark effect of the excitons in semiconductors 
\cite{dyn} in a wide range of semiconductor bulk and quantum well systems.
With recent advances in ultrafast nonlinear optical spectroscopy and in
fabrication of semiconductor heterostructures,
the use of three-pulse FWM, polarization \cite{fink,benn1} and phase sensitive
measurements of the nonlinear polarization \cite{chem}, and 
non-degenerate transient FWM \cite{cund} has led to effects beyond the mean-field
approximation for exciton interaction and beyond the Markovian approximation
for dephasing. Effects such as the polarization-dependent response
of the excitons \cite{benn1,exp} and signatures of bound biexcitons 
\cite{love,may2,bart1} lead to more
refined theoretical investigations beyond the mean-field approximation
\cite{fink,benn1,fwm,axt,may,diag1,lind,axt2,th1,schaf2,kner}.


Polarization-mixing in the nonlinear optical response is caused by excitation of
two excitons of opposite spins which first occur in second order
of the Coulomb interaction \cite{diag1}.  The mean-field description
which treats the interaction between excitons to first order in the Coulomb
interaction can account for neither the polarization mixing nor the formation of
bound biexcitons. This point was clearly demonstrated by Combescot and
Combescot \cite{c2}  who had stressed the importance of bound and unbound
biexciton states for the excitonic ac-Stark shift as well as polarization
effects \cite{mcom}.
The formation of para-biexcitons with singlet spin-states for the electrons
and holes is one aspect of polarization-mixing which is most likely to be
dominant for near resonant excitations of the fundamental exciton states.
Moreover, even in the absence of bound biexcitons, the correlation in the
continuum of two-exciton scattering states is also important \cite{exp,th1}.


In studying the optical processes in semiconductor systems,  while much can be
learned from ensembles of non-interacting atomic transitions
\cite{allen,agar,ct} interacting with the radiation field or 
interacting localized (dense) two-level  systems \cite{muka}, the strong
interaction and close proximity of the electrons in semiconductors provides a
distinct avenue for new physics.  It is well known that many-body effects
lead to a renormalization of the external field (local-field effects) as well as
to a renormalization of the interband transition energies (self-energy effects),
depending sensitively on the density and dynamics of the surrounding
electron-hole pairs \cite{sr,srrev}. In the linear response regime, these
effects assume quantitative importance \cite{hanke}.
However, the nonlinear properties of semiconductors in the weak nonlinear
regime and in the high-excitation regime constitute 
challenging problem, where the electron interaction physics must be added to
carrier non-equilibrium and physics of quantum optics.


In the low-density or weakly nonlinear regime, i.e., to third-order in the
external field, the dynamics of the semiconductor for near resonant excitation
of the fundamental  exciton resonances can be formulated in terms of a set of
effective dynamical equations for the exciton polarization with nonlinear
exciton-exciton interaction and space filling effects, which have been derived
from the semiconductor Bloch equations (SBE) \cite{exho,kuk,non}. These effective
equations provide a useful and physical picture of the origin of the
nonlinearities and  the observed phenomena, for example
in the theory of the four-wave-mixing \cite{exho}, 
photon echo \cite{echo} or the Rabi-oscillations in semiconductors \cite{rabi1}. 
Being within a mean-field description, the effective equations can provide
no more information than the full semiconductor Bloch equations
in the mean-field approximation.


The inclusion of biexcitonic effects as well as the possibility
of polarization-mixing was first discussed in terms of phenomenological 
few-level models \cite{fink,benn1,fwm} to explain the temporal dependence of the
FWM-signal signal including oscillations as a beating phenomena between bound and
unbound biexciton states.  In more microscopic theories, 
it was first shown in the equation-of-motion method by Axt and Stahl \cite{axt}
and later in a diagrammatic approach by Maialle and Sham \cite{diag1} that
the semiconductor Bloch equations form a closed set of equations
for the density matrix elements for any given order of the external field,
depending on the initial state, which is usually the vacuum state 
of the electron-hole pairs.
The so-called  dynamics-controlled truncation scheme \cite{axt}
provides a starting point for a microscopic theory of the polarization effects
 and has been applied to up to fifth-order processes \cite{bart1}.
The inclusion of biexcitonic effects for actual applications
is mostly treated in a
restriction to bound-state contributions and a {\it single}
two-exciton state contribution \cite{axt2} or
perturbation theory in the Coulomb interaction \cite{diag1,lind}.
The solution for the third-order susceptibility is inextricably bound to the
solution of the four-particle problem.


In this paper, we give a microscopic theory of exciton interaction
effects in nonlinear optical processes based on the Coulomb interaction between
electrons.  The detailed account provides the derivation behind our 
results published earlier \cite{th1,lu}.  The theory recovers the established
mean-field results in the literature and formulates the rest of the
interaction effects, termed correlation, in a concise manner. Particularly
striking is the resulting equation of motion for the third-order nonlinear
optical response shown to be driven by a number of terms with clearly identified
physical origins: (1) phase-space-filling corrections, which are due to the
Pauli-blocking of electrons and holes, (2) exciton-exciton mean-field
interaction, and (3) the correlation term which is expressed as a 
two-exciton force-force correlation function.  The derivation of the general
equations of motion for any given order of the external field is given in
Sec.~\ref{section2} in terms of the Hubbard operators, using a complete basis 
set of the $N$ eh-pair states.  The Axt-Stahl theorem \cite{axt}  manifests
itself as, for example, the Hilbert space of one and two electron-hole pairs
being sufficient for the third-order nonlinear response of the semiconductor.  
Details of the commutation algebra are relegated to Appendix \ref{apena}.  An
alternative derivation in terms of the density matrix is not recorded here to
keep the length of the paper within bounds.  
The part of
dephasing which is due to the electron interaction effects is included in our
correlation function and the rest of the dephasing due to other causes is
treated phenomenologically. 


The correlation function approach gives a unified description of all
correlation effects.  It naturally encompasses the recently observed
polarization mixing and bound-state biexcitonic molecules.  The exact
two-exciton correlation treats these effects on the same footing as the
two-exciton scattering states which will be shown to be equally important.  It
contains the exact spin-dependent Coulomb correlation among the four
particles and determines the spectral weight of the biexciton states for the
source-term of the nonlinear response. These properties can be demonstrated by
a numerical example of a one-dimensional model system with the advantage of not
making decoupling approximations of the correlation effects.  (See 
Sec.~\ref{section3}).  Details of the model are described in Appendix
\ref{apenb}.

Section \ref{section4} gives an example of the correlation effects on a
nonlinear optical process: a three-pulse four-wave-mixing experiment. 
Extensive new results of the numerical evaluation of the simple model are
qualitatively compared with experimental results. Exact numerical evaluation of
the correlation function is confined to simple models.  For more realistic
models to the semiconductor systems, we need reasonable approximations. A
number of these are investigated in section \ref{section5}, including a 
microscopic expression for the excitation induced dephasing \cite{wang2} and a
brief comparison with the Boltzmann corrections in the quantum-kinetic 
equations \cite{tran1}.  We conclude with a summary of our theory and with a
brief outline of future applications in section \ref{section6}.




%
\section{An equation of motion for the third-order response}
\label{section2}
%

We take as the fundamental approximation that, in the absence of the light-matter
interaction, the Hilbert space of the semiconductor model consists of
disconnected subspaces, which can be labeled according to the number of 
electron-hole (eh) pairs in the many-particle states. Let $|0\rangle $ denote 
the trivial ground state with no eh-pairs present with energy $\omega_{0,0}=0$. 
The one eh-pair subspace is the exciton subspace with states
$|E^{(1)}_{n,\sigma}\rangle$ with the quantum number $n$, a
polarization index $\sigma$ and energy $\omega_{1,n,\sigma}$. 
Both bound and scattering states are included in $n$.
The polarization
index  labels a specific transition which, if optically active, corresponds to
the helicity of the light  required to excite the 
eh-pair-states that form the exciton. For example, in Zincblende structures,
four p-type valence band states with  
total angular momentum $3/2$ are connected via an optical dipole transition
to an s-type conduction band with spin degeneracy. Due to selection rules, the
$m=3/2 (m=-3/2)$ electrons in the heavy-hole band 
are coupled via an optical transition with $-(+)$ 
polarized photons to the $s=1/2 (s=-1/2)$ spin states in the conduction
band.  The $m=1/2 (m=-1/2)$ electrons in the light-hole band 
are coupled via an optical transition with $-(+)$ 
polarized photons to the $s=-1/2 (s=1/2)$ spin states in the conduction
band. The spin-orbit interaction usually splits off a valence-band
with total angular-momentum $1/2$ and is neglected throughout this
investigation.

The next relevant subspace is the biexciton Hilbert space
with a complete set $|E^{(2)}_{m}\rangle$ of bound and unbound
states. Here, 
we introduce a single index $m$ to label the set of quantum numbers for 
the states with energy $\omega_{2,m}$.  Even though not all the biexciton states
are computed due to the many-body nature of the problem, we keep all the states
as long as possible because occasions arise that the biexciton states as
intermediate states can be re-summed by virtue of the completeness theorem,
similar to the treatment of the  ac-Stark shift \cite{c2}.  Such a step would
be lost in a common approximation which restricts from the start to one or two
biexciton states.

The use of the subspaces of different exciton numbers as disconnected is
implicit in previous works \cite{axt,diag1}.  The disconnectedness is an
approximation because states with different exciton numbers can be connected by
the Coulomb interaction.  The most important consequence is
the neglect of the electron-hole pair fluctuations which affects the ground
state and the dielectric screening of the Coulomb interaction.  In other words,
we define the ground state of the semiconductor as a vacuum state with respect 
to the exciton annihilation
\begin{eqnarray}
B_{n,\sigma} |0 \rangle = 0, \label{gs}
\end{eqnarray}
which means that no electron-hole pairs are present in the
semiconductor ground state.  The dielectric screening is approximately
accounted for by the static dielectric constant of the semiconductor.

We define a total $\sigma$-polarization connected to an optical transition,
which can be of arbitrary helicity, depending on the electronic states
involved,
\begin{eqnarray}
P_\sigma = \mu^\ast_\sigma  \sum_{\bf k}  
\psi_{{\bf k},\sigma}
\end{eqnarray}
where the operator $\psi^\dagger_{{\bf k},\sigma}$ 
creates a zero total momentum
eh-pair with electron wave-vector ${\bf k}$, hole wave-vector $-{\bf k}$ and
polarization index $\sigma$ and
$\mu_\sigma$ is the dipole matrix element between the electron and hole states,
assumed to be independent of ${\bf k}$. Completeness of the operators leads to
an equivalent expression of the
$\sigma$-polarization in terms of  exciton creation operators:
\begin{eqnarray}
 P_\sigma =
 \mu^\ast_\sigma \sum_n \alpha_{n,\sigma} B_{n,\sigma} \; ,
\label{pola} 
\end{eqnarray}
where
\begin{eqnarray}
\alpha_{n,\sigma}
= \sqrt{V}  \Phi_{n,\sigma}({\bf x}=0),    \label{alpha}
\end{eqnarray}
in terms of the exciton wave function at zero relative distance. 
The operator 
\begin{eqnarray}
B_{n,\sigma} = \sum_{\bf k} \psi_{{\bf k}, \sigma}
\phi_{{\bf k},n,\sigma}^\ast
\end{eqnarray}
creates an exciton state 
$B_{n,\sigma}^\dagger|0\rangle =|E^{(1)}_{n,\sigma}\rangle $
with zero total momentum, energy $\omega_{1,n,\sigma}$
and relative wave function $\phi_{{\bf k},n,\sigma}$ 
in terms of momentum 
${\bf k}$.  The combination of the electron band $\lambda_1$ and the hole band
$\lambda_2$ determines the polarization index
$\sigma=\sigma(\lambda_1,\lambda_2)$.
The laser central frequency $\omega_p$ is implicitly subtracted from the exciton
energy when we transfer to the rotating frame. The biexciton energy
$\omega_{2,m}$ then contains a reduction $-2\omega_p$.   Note that the factor
$\alpha_{n,\sigma}$ depends on the sample volume $V$ of the system for {\it
bound} exciton states and is nonzero only for exciton states with $s$-wave
symmetry. We take care of this volume dependence and show clearly how the final
result is indeed volume independent.

Using the Dirac notation, we introduce the following Hubbard operators
\begin{eqnarray}
\hat{X}_{N,\alpha;M,\beta} = |E_{N,\alpha}\rangle \langle E_{M,\beta}|
\label{hub}
\end{eqnarray}
which can be used, in combination with the completeness relation
\begin{eqnarray}
  {\bf 1}  \equiv \sum_{N,\alpha} \hat{X}_{N,\alpha;N,\alpha} \; ,
  \label{id}
\end{eqnarray}
to express the exciton operator as
\begin{eqnarray}
B_{n,\sigma} &=& \hat{X}_{0;1,n,\sigma}
+  \sum_{N\ge1,\alpha,\beta} 
 \langle E_{N,\alpha} | B_{n,\sigma}| E_{N+1,\beta} \rangle
\hat{X}_{N,\alpha;N+1,\beta}.
\end{eqnarray}
The interaction of the semiconductor with a classical external laser field
with central frequency $\omega_p$ and  field-strength
${ \bf E}(t)=\sum_\sigma E_\sigma(t) e^{-i\omega_p t} {\bf e}_\sigma
 + c.c.$ is given
in the usual rotating wave approximation \cite{allen} by 
%
%
\begin{eqnarray}
H_I &=& -  \sum_{n,\sigma}
\left(E_{n,\sigma}^\ast(t) B_{n,\sigma} 
 +
{\rm h. c.} \right) 
\end{eqnarray}
with 
\begin{eqnarray}
E_{n,\sigma}(t)=\mu_\sigma \alpha_{n,\sigma}^\ast  E_\sigma(t) \label{rabi}
\end{eqnarray}
For comparison with previous work \cite{non}, this expression
is the time-dependent renormalized 
Rabi frequency of a given polarization $\sigma$ and transition $n$
($\hbar\equiv 1$).
The Hamiltonian of the semiconductor, from the disconnectedness of the
subspaces, is
\begin{eqnarray}
H = \sum_{N,\alpha} \omega_{N,\alpha} \hat{X}_{N,\alpha;N,\alpha} 
\end{eqnarray}
which is equivalent to a multi-band microscopic Hamiltonian
in second quantization.
From the form of the interaction $H_I$ it follows that
the expectation values 
$ \langle \hat{X}_{0;N,\alpha} \rangle_t$
can be expressed as a power series in the external field
\begin{eqnarray}
 \langle \hat{X}_{0;N,\alpha}\rangle_t &=& 
\sum_{m=0}^{m_0} X^{(N+2m)}_{N,\alpha}(t)
+ O(E^{N+2m_0+2})  \label{w4} \;.
\end{eqnarray}
The expectation value of a zero to $N$-pair transition is at least of
order $N$ in the external field. This {\it theorem} has already been proven
by Axt and Stahl \cite{axt}.
%
%
%
%
An important relation can be derived by the identity
for an arbitrary state $|\phi(t)\rangle$
\begin{eqnarray}
 \langle \hat{X}_{N,\alpha; M,\beta}\rangle_t &=& 
 \langle \phi(t)| \hat{X}_{N,\alpha;M,\beta}
 | \phi(t) \rangle \nonumber  \\
&=&  \langle \phi(t) |E_{N,\alpha}\rangle \langle 0|  \phi(t)
  \rangle 
 \langle \phi(t)| 0\rangle \langle E_{M,\beta}| \phi(t) \rangle
\langle \hat{X}_{0;0} \rangle_t^{-1} \nonumber .
\end{eqnarray}
With  $ \hat{X}_{0}^{(0)}(t) =1$
from  the initial condition of the semiconductor
in its ground state $|0 \rangle$, we find 
for the general expectation values
\begin{eqnarray}
 \langle \hat{X}_{N,\alpha;M,\beta}\rangle_t &=& 
 \langle \hat{X}_{0;N,\alpha}\rangle_t^\ast
 \langle \hat{X}_{0;M,\beta}\rangle_t \langle
 \hat{X}_{0;0}\rangle_t^{-1}.   \label{w5}
\end{eqnarray}
%
%

In order to calculate the $\sigma$-polarization we consider the equation of
motion for $\langle B_{n,\sigma}\rangle_t$. Using the Hubbard operators
it reads
\begin{eqnarray}
i\frac{\partial}{\partial t} 
\langle B_{n,\sigma}\rangle_t
& = & (\omega_{1,n,\sigma} -i\Gamma) \langle B_{n,\sigma}\rangle_t
  + \sum_{N\ge 1} 
\sum_{\alpha,\beta}
 c_{n,\sigma;\alpha,\beta}^{(N)} \langle 
\hat{X}_{N,\alpha;N+1,\beta} \rangle_t 
\label{finalp} 
\\ &-& E_{n,\sigma}(t)
 + \sum_{N\ge 1}\sum_{\alpha,\beta}
 E_{n,\sigma;\alpha,\beta}^{(N)} \langle 
\hat{X}_{N,\alpha;N,\beta} \rangle_t  \nonumber
\end{eqnarray}
with
\begin{eqnarray}
 c_{n,\sigma;\alpha,\beta}^{(N)} &=& (\omega_{N+1,\beta}-\omega_{N,\alpha}
 -\omega_{1,n,\sigma})
 \langle E_{N,\alpha} | B_{n,\sigma}| E_{N+1,\beta} \rangle
\label{pre1}\\
 E_{n,\sigma;\alpha,\beta}^{(N)} &=&
 \sum_{\bf q} \mu_{\sigma} E_{\sigma}(t)
\phi_{{\bf q},n,\sigma} ^\ast
 \langle  
 E_{N,\alpha} | 1 - n_{1,{\bf q},\sigma} + n_{2,{\bf q},\sigma}
| E_{N,\beta} \rangle    \label{pre2} .
\end{eqnarray}
We have introduced the dephasing due to degrees of freedom not
included explicitly (e.g. phonons) in a phenomenological way
with the effective parameter $\Gamma$. 
Using Eqs.~(\ref{w4}) and (\ref{w5}) we see that Eq.~(\ref{finalp})
can be considered as a linear differential equation with a (trivial)
first-order source  and nontrivial source terms
of third and higher order. 
In the following we restrict
ourselves to the contributions up to {\it third} order. Then using
Eqs.~(\ref{w4}) and (\ref{w5}) we see that only $X^{(1)}_{1,n,\sigma}$
and $X^{(2)}_{2,\alpha}$ have to be determined.
As  $X^{(1)}_{1,n,\sigma}(t)$ obeys Eq.~(\ref{finalp}) without the
terms involving the summations one obtains
\begin{eqnarray}
 X_{1,n,\sigma}^{(1)}(t) &=&  i
  \int_{-\infty}^t 
e^{-i(\omega_{1,n,\sigma}-i\Gamma)(t-t')} 
 E_{n,\sigma}(t') dt' \;.  \label{linpola}
\end{eqnarray}
The equation of motion for $X^{(2)}_{2,\beta}$ reads
\begin{eqnarray}
i \frac{\partial}{\partial t} X_{2,\beta}^{(2)} &=& 
  \left( \omega_{2,\beta} -i\Gamma_{xx}
\right) X_{2,\beta}^{(2)} \label{B}  \\
&-&  \sum_{n',\sigma'; n^{''},\sigma^{''}}
E_{n^{''},\sigma^{''}}(t)
 \langle E_{2,\beta} | B_{n^{''},\sigma^{''}}^\dagger
| E_{1,n',\sigma'} \rangle
X^{(1)}_{1,n',\sigma'}   \nonumber
\end{eqnarray}
with the biexciton
 phenomenological dephasing
constant $\Gamma_{xx}$.
In order to write the second term on the rhs of Eq.~(\ref{finalp})
in a compact form we use the explicit result Eq.~(\ref{B}) for
$X^{(2)}_{2,\beta}(t)$ in order to perform the summation
over the biexciton quantum numbers $\beta$. 
\begin{eqnarray}
X_{2,\beta}^{(2)}(t) &=& 
  i
\sum_{n',\sigma'; n^{''},\sigma^{''}}
\int_{-\infty}^t e^{-i\left(\omega_{2,\beta} -i \Gamma_{xx}
\right)(t-t')} \nonumber \\ &\times &
E_{n^{''},\sigma^{''}}(t')
 \langle E_{2,\beta} | B_{n^{''},\sigma^{''}}^\dagger
| E_{1,n',\sigma'} \rangle
X^{(1)}_{1,n',\sigma'}(t')  
\end{eqnarray}
Using the identity
\begin{eqnarray}
&&  i
\sum_{n',\sigma'; n^{''},\sigma^{''}}
E_{n^{''},\sigma^{''}}(t)
X^{(1)}_{1,n',\sigma'}(t)  
e^{i\left(\omega_{1,n'',\sigma''} + \omega_{1,n',\sigma'} -2i\Gamma \right)t} 
\nonumber \\ 
&& = \frac{1}{2} \partial_t \left( \rule[-3mm]{0cm}{0.6cm}
\sum_{n',\sigma'; n^{''},\sigma^{''}}
X^{(1)}_{n^{''},\sigma^{''}}(t)
X^{(1)}_{1,n',\sigma'}(t)  
e^{i\left(\omega_{1,n'',\sigma''} + \omega_{1,n',\sigma'} -2i\Gamma
  \right)t} \right)
\end{eqnarray}
we obtain after a partial integration
\begin{eqnarray}
 &&  \sum_{\beta}
  c_{n,\sigma;\tilde{n},\tilde{\sigma},\beta}^{(1)} 
  X_{2,\beta}^{(2)}(t)    \label{bdyn1}
\\ &=&
  \frac{1}{2} \sum_{n',\sigma',n^{''},\sigma^{''}} \nonumber 
  \left\{ \rule[-5mm]{0cm}{1cm}
    \langle E_{1,\tilde{n},\tilde{\sigma}} |
    B_{n,\sigma} \left( H - \omega_{1,n,\sigma} 
      - \omega_{1,\tilde{n},\tilde{\sigma}} \right)
    B^\dagger_{n^{''},\sigma^{''}}
    |E_{1,n',\sigma} \rangle \right. \nonumber \\ &\times&
  X^{(1)}_{1,n',\sigma'}(t) X^{(1)}_{1,n^{''},\sigma^{''}}(t) 
  \nonumber \\
  &-& \sum_\beta
  \int_{-\infty}^t 
  \partial_{t'} \left( \rule[-3mm]{0cm}{0.6cm}
    \langle E_{1,\tilde{n},\tilde{\sigma}} |
    B_{n,\sigma} 
e^{-i(\omega_{2,\beta} -i \Gamma_{xx})(t-t')}
    |E_{2,\beta} \rangle \right. \nonumber \\
&\times& \left.
    \langle E_{2, \beta} |
    \left( H - \omega_{1,n,\sigma} 
      - \omega_{1,\tilde{n},\tilde{\sigma}} \right)
    B^\dagger_{n^{''},\sigma^{''}}
    |E_{1,n',\sigma} \rangle 
    e^{i\left(\omega_{1,n'',\sigma''} + 
        \omega_{1,n',\sigma'} -2i\Gamma \right)(t-t')}
\rule[-3mm]{0cm}{0.6cm} \right) 
    \nonumber \\
    &\times&  \left.
    X^{(1)}_{1,n',\sigma'}(t') X^{(1)}_{1,n^{''},\sigma^{''}}(t')
      e^{-i\left(\omega_{1,n'',\sigma''} + \omega_{1,n',\sigma'} -2i\Gamma
        \right)(t-t')}
     dt' \rule[-5mm]{0cm}{1cm}
\right\} \nonumber .
\end{eqnarray}
We can also perform  the $\beta$-summation in the second term on the rhs of 
Eq.~(\ref{bdyn1}). For the derivative, we use the following identity,
which holds for 
$[B_{\tilde{n},\tilde{\sigma}},B_{n,\sigma}]=0$ and $H|0\rangle =0$
\begin{eqnarray}
&  \partial_\tau & \left( \rule[-3mm]{0cm}{0.6cm}
    \langle E_{1,\tilde{n},\tilde{\sigma}} |
    B_{n,\sigma} \left( H - \omega_{1,n,\sigma} 
      - \omega_{1,\tilde{n},\tilde{\sigma}} \right)
    e^{-iH\tau } B^\dagger_{n^{''},\sigma^{''}}
    |E_{1,n',\sigma} 
\label{id2}
\rangle \right. \\ &\times& \left. e^{i\left(\omega_{1,n'',\sigma''} + 
        \omega_{1,n',\sigma'} \right)\tau} \rule[-3mm]{0cm}{0.6cm} \right) 
\nonumber \\ 
&\equiv &
 -i \langle 0| D_{\tilde{n},\tilde{\sigma};n,\sigma}(\tau) 
D_{n',\sigma';n^{''},\sigma^{''}}^\dagger 
|0 \rangle     
\; e^{i\left(\omega_{1,n'',\sigma''} + 
        \omega_{1,n',\sigma'} \right)\tau} \;.
\nonumber
\end{eqnarray}
Here we have introduced the ``force'' operator
\begin{eqnarray}
  D_{\tilde{n},\tilde{\sigma};n,\sigma}
  = [B_{\tilde{n},\tilde{\sigma}},[B_{n,\sigma},H]] \;
\end{eqnarray}
and the usual time dependence in the Heisenberg picture
is given by $D(\tau)=e^{iH\tau}De^{-iH\tau}$.
This allows to write Eq.~(\ref{bdyn1}) in a compact form
defining a memory kernel
\begin{eqnarray}
  F_{\tilde{n},\tilde{\sigma};n,\sigma}^{n',\sigma';n^{''},\sigma^{''}}
  (\tau) &:=& 
  \langle 0| D_{\tilde{n},\tilde{\sigma};n,\sigma}(\tau) 
  D_{n',\sigma';n^{''},\sigma^{''}}^\dagger 
  |0 \rangle  \;,    \label{force}
\end{eqnarray}
for the second term on the rhs of Eq.~(\ref{bdyn1}). The
matrix element in the mean-field contribution (first term on the rhs
of Eq.~(\ref{bdyn1})) can also be simplified,
which gives
\begin{eqnarray}
  \langle E_{1,\tilde{n},\tilde{\sigma}} |
  B_{n,\sigma} \left( H - \omega_{1,n,\sigma} 
    - \omega_{1,\tilde{n},\tilde{\sigma}} \right)
&=&
  \langle E_{1,\tilde{n},\tilde{\sigma}} |
  \left( [B_{n,\sigma}, H] - \omega_{1,n,\sigma} \right)
\\ &=&
  \langle 0| D_{\tilde{n},\tilde{\sigma};n,\sigma} \nonumber
\end{eqnarray}
and we finally arrive at
\begin{eqnarray}
  \sum_{\beta}
  c_{n,\sigma;\tilde{n},\tilde{\sigma},\beta}^{(1)} 
  {X}_{2,\beta}^{(2)}(t)  &=&
  \label{bdync}
  \frac{1}{2} \sum_{n',\sigma',n^{''},\sigma^{''}} 
  \left\{ \rule[-3mm]{0cm}{0.6cm} 
    \langle 0| D_{\tilde{n},\tilde{\sigma};n,\sigma}
    B^\dagger_{n',\sigma'} B^\dagger_{n^{''},\sigma^{''}}
    |0\rangle \right. \\ &\times&
  X^{(1)}_{1,n',\sigma'}(t) X^{(1)}_{1,n^{''},\sigma^{''}}(t) 
  \nonumber \\
  &-& 
  i
  \int_{-\infty}^t 
  e^{-2\Gamma(t-t')} 
  F_{\tilde{n},\tilde{\sigma};n,\sigma}^{n',\sigma';
    n^{''},\sigma^{''}}
  (t-t')  \nonumber \\ &\times& \left. 
    X^{(1)}_{1,n',\sigma'}(t') X^{(1)}_{1,n^{''},\sigma^{''}}
    (t') dt' \rule[-3mm]{0cm}{0.6cm} \right\} .
  \nonumber
\end{eqnarray}
Here it is necessary to assume the relation $\Gamma_{xx} \equiv
2\Gamma$
to use the identity Eq.~(\ref{id2}).
The memory function in Eq.~(\ref{force}) is 
a four-point correlation function in terms of electron (hole) operators.
As the operators
$A_{n,\sigma}({\bf q})$
can be expressed in terms of {\it finite}
center-of-mass exciton operators $B_{n,\sigma}({\bf q})
$, $F(\tau)$ can be considered a two-exciton 
correlation function. From the double commutator definition of the $D$
operators, Eq.~(\ref{force})
is a {\it force-force} correlation function \cite{mahan}.
An explicit form of the operator $D$ is derived in Appendix A using a
 specific semiconductor model.

The first expression on the rhs of Eq.~(\ref{bdync})
describes the correlations between excitons 
as in the usual mean-field semiconductor Bloch equations (MFSBE)
\cite{koch88,exho,non,echo}.
It is only nonzero for excitons with zero center-of-mass momentum
and with {\it identical polarization}, i.e., each pair of charged
carriers
(electrons and holes)
must belong to the same bands,
and consequently does not produce polarization mixing.
For comparison with previous work on the 
MFSBE and earlier work on exciton-exciton interaction
\cite{51,hana,zimbu}, we introduce the matrix elements
\begin{eqnarray}
  \beta_{\tilde{n},\tilde{\sigma};n,\sigma}^{n',\sigma';
    n^{''},\sigma^{''}} &=&
    \langle 0| D_{\tilde{n},\tilde{\sigma};n,\sigma}
    B^\dagger_{n',\sigma'} B^\dagger_{n^{''},\sigma^{''}}
    |0 \rangle, 
\label{mfpara}
\\ 
  \gamma_{\tilde{n},\tilde{\sigma};n,\sigma}^{n',\sigma';
    n^{''},\sigma^{''}} &=&
  \langle 0| D_{\tilde{n},\tilde{\sigma};n,\sigma}
  D^\dagger_{n',\sigma';n^{''},\sigma^{''}}
  |0 \rangle \;.  \label{mfpara2}
\end{eqnarray}
As 
$D_{\tilde{n},\tilde{\sigma};n,\sigma}(\tau)|0\rangle = 0$,
the correlation function $F(\tau)$
can be written as a time-ordered product and standard Feynman diagrams
can be used e.g. to set up approximation schemes. From a diagrammatic analysis
to all orders the rigorous polarization selection rule 
$\sigma+\tilde{\sigma}=\sigma'+\sigma''$
can easily be read off \cite{diag1}.
The fact that the third-order polarizability can be expressed in terms of this
correlation function depending on a {\it single} time difference
is due to the simplicity of the semiconductor ground state Eq.~(\ref{gs})
approximated by the vacuum state of the bound and unbound excitons.

An additional contribution to the third-order nonlinear response is given by
the phase-space filling factor, which is due to the Pauli blocking 
of electrons. This term is assumed to play a minor role in the low-density
regime of optical excitations of semiconductors, but we include this
contribution here to preserve the exactness of our expression to third order in
the exciting field. Keeping the $N=1$ contribution in Eq.~(\ref{pre2}) we find
with Eq.~(\ref{w5})
\begin{eqnarray}
 \sum_{\alpha,\beta}
 E_{n,\sigma;\alpha,\beta}^{(1)} \langle 
\hat{X}_{1,\alpha;1,\beta} \rangle_t &=&
\sum_{\tilde{n},\tilde{\sigma}, n',\sigma', n^{''},\sigma^{''}} 
 C_{\tilde{n},\tilde{\sigma};n,\sigma}^{n',\sigma'; 
    n^{''},\sigma^{''}}
  E_{n',\sigma'}(t) \\ &\times&
    \left( X^{(1)}_{1,\tilde{n},\tilde{\sigma}}(t)\right)^\ast
  X^{(1)}_{1,n^{''},\sigma^{''}}(t)   \nonumber
\end{eqnarray}
The phase-space-filling parameter depends on the explicit
exciton wave-functions, c.f. Appendix~\ref{apena}:
\begin{eqnarray}
 C_{\tilde{n},\tilde{\sigma};n,\sigma}^{n',\sigma'; 
    n^{''},\sigma^{''}} &=&
 \delta_{\sigma,\sigma'} \alpha_{n',\sigma'}
\sum_{{\bf q}} 
 \phi_{{\bf q},n,\sigma}^\ast 
\nonumber \\ &\times&
 \langle  
 E_{1,\tilde{n},\tilde{\sigma}} | 1 - n_{1,{\bf q},\sigma} + n_{2,{\bf q},\sigma}
| E_{1,n'',\sigma''} \rangle    \label{pauli} .
\end{eqnarray}
The exact third-order nonlinear polarization $P_{n,\sigma}^{(3)}(t)$
is given by the solution of the following
linear differential-equation with a complete set of source terms
\begin{eqnarray}
  & & \left(\frac{\partial}{\partial t}
    +i \omega_{1,n,\sigma} + \Gamma \right)  P^{(3)}_{n,\sigma}(t)
  = -i  
  \sum_{\tilde{n}, \tilde{\sigma}, n',\sigma', n^{''},\sigma^{''}} 
  \left( X^{(1)}_{1,\tilde{n},\tilde{\sigma}}(t)\right)^\ast
  \label{pola_result}
  \\ &  \left\{ \rule[-3mm]{0cm}{0.6cm} \right. & 
  C_{\tilde{n},\tilde{\sigma};n,\sigma}^{n',\sigma'; 
    n^{''},\sigma^{''}}
  E_{n',\sigma'}(t)
  X^{(1)}_{1,n^{''},\sigma^{''}}(t) 
  +
  \frac{1}{2} 
    \beta_{\tilde{n},\tilde{\sigma};n,\sigma}^{n',\sigma'; 
    n^{''},\sigma^{''}}
  X^{(1)}_{1,n',\sigma'}(t) X^{(1)}_{1,n^{''},\sigma^{''}}(t) 
  \nonumber \\
  &-& 
    \frac{i}{2}
    \int_{-\infty}^t 
    e^{-2\Gamma(t-t')} 
    F_{\tilde{n},\tilde{\sigma};n,\sigma}^{n',\sigma';
      n^{''},\sigma^{''}}
    (t-t')  
    X^{(1)}_{1,n',\sigma'}(t') X^{(1)}_{1,n^{''},\sigma^{''}}
    (t') dt' 
    \nonumber   \left. \rule[-3mm]{0cm}{0.6cm} \right\}.
\end{eqnarray}
This equation expresses succinctly the physical origins of the source terms which
drive the third-order polarization: the first term in the curly bracket being the
phase-space filling, the second the Hartree-Fock or mean-field terms (of first
order in the Coulomb interaction between excitons), and the last the biexciton
correlation. The source terms require a solution of the linear-response problem
of Eq.~(\ref{linpola}). The above derivation of the third-order nonlinear
response can be extended to higher order in the external field. However, the
higher-order correlation functions involved  no longer have the simple
structure of the third-order response force-force correlation function.






       
%
\section{Exciton-Exciton correlations}
\label{section3}
%

The correlation function formulation of the last section provides us with a
powerful framework to compute the correlation effects of the two excitons
based on treating the interaction among the two electrons and two holes on a
equal basis.  In this section, we start with the nonlinear optical
processes which lead to the excitation of the excitons and discuss some general
properties inferred from a study of a one-dimensional system, where the
correlation function can be calculated numerically for relatively large system
sizes without approximating the effects of the many-body interaction.  
 We calculate the spectral-function for the
$1s$-exciton contribution to the third-order optical response for a
one-dimensional semiconductor model. This is done in the frequency representation
with the use of the Lanczos algorithm \cite{lanz1}. The details of the 
one-dimensional semiconductor model in real-space are defined in
Appendix~\ref{apenb}. 

The selection rule which connects the helicity of the the exciting light to the
spins of the electron and hole gives rise to different types of two-exciton
excitations depending on the energy level structures. For a semiconductor with 
Zincblende structure, where spin-orbit leads to a four-fold degenerate
valence-band maximum, Fig.~\ref{fig4} illustrates the three types: 
\begin{enumerate}

\item[\it Type-I]
  
  In heterostructures, the strain or confinement
  can lead to a considerable heavy- and light-hole splitting
  $\Delta_{hl-lh}$. This allows for a selective excitation of heavy-
  or light-hole exciton states.

\item[\it Type-II]
  
  In bulk systems,  circularly polarized light
  excites a combination of  heavy- and light-hole excitons simultaneously.

\item[\it Type-III]
  
  Polarization mixing can be induced by a linearly polarized light.
  A special situation arises, when conduction band states are common
  for both excited exciton species, in contrast to the type-II excitation.
  A three-pulse four-wave-mixing geometry can distinguish this analog of the
  {\it $\Lambda$-transition} in quantum optics from the decoupled transitions.

\end{enumerate}


Correlation among the electrons and holes depends sensitively on their 
angular momenta (spins) and, in turn, influences the polarization dependence in
the nonlinear optical response. We sort out for each combination of spins the
possible types of excitations listed above:

\begin{itemize}

\item {\it Opposite-spin excitons:} 
  
  If the two excitonic transitions belong to different conduction and
  valence bands, i.e., no single-particle states are the same, e.g., 
                the polarization mixing of simultaneous excitations of the
   $m=-3/2$ valence band to conduction-band $s=-1/2$ transition
  with a {\it positive}
  circular polarization of the light-field and 
  the $m=3/2$ valence band to conduction-band $s=1/2$ states with a 
  {\it negative}
  circular polarization of the light-field respectively,
  this excitation condition is type-I.
  In the absence of a significant band-splitting, e.g., in bulk
  systems, a {\it positive} circular excitation will excite light-hole
  excitons from valence band $m=-1/2$ to conduction-band $s=1/2$ states
  as well as from 
  valence band $m=-3/2$ to conduction-band $s=-1/2$.
  This situation belongs to the type-II excitation. Although both
  excitons have equal-spin, the exciton energies and transition strengths are
  different and can be distinguished.
  In both cases above, (one type-I and one type-II), the hole {\it
  and} the
  electron states belong to different bands. The correlation among
  the four charged carriers is now entirely determined by Coulomb
  effects since no Pauli-blocking is encountered in the band.

\item {\it Parallel-spin excitons:} 

  When the two exciton transitions have common hole and electron bands,
  their spins are parallel. For a type-I
  excitation, with only, say, positive circularly polarized excitation,
  only spin $+1$-excitons are populated. A three-pulse experiment can 
  distinguish between the equal-spin and the opposite-spin correlations
  in the response signal. This point is discussed in the context of the
  four-wave-mixing experiments in the next section.
  The most prominent feature of the equal-spin
  correlation is the absence of a bound biexcitonic molecule. This shows
  up clearly  in the one-dimensional model.
  A type-II excitation will mix the opposite- and parallel-spin
  correlations
  and the analysis of the FWM geometry is more complicated, due to the
  natural interference of the individual contributions.
  
\item {\it Coupled-spin excitons:} 

  The type-III excitation arises when either the electron or hole states
  but not both share a common band.  For example, linear excitation of heavy-
                and light-hole excitons in $GaAs$ will induce transitions from the 
  $m=-3/2$ valence states with photons of positive helicity as well as
  transitions
  from the $m=1/2$ valence states with photons of negative
  helicity to common conduction-band states with $s=-1/2$. 
  The corresponding correlation function is similar to the parallel-spin case
  in the sense that the conduction-band electrons are subject to the
  Pauli exclusion.  With a finite heavy/light hole splitting, this type-III
  excitation is sensitive to the time delay between the two exciting pulses 
  in a three-pulse FWM experiment and can, thus, be distinguished from the
  parallel-spin type-I case.
  
\end{itemize}


Fig.~\ref{fig1} shows the correlation function spectrum for the
diagonal $1s$-exciton con\-tribution calculated in
a one-dimensional semiconductor model. The oppo\-site-spin case (solid line)
and the parallel-spin case (dashed line) are  shown. The zero of
energy corresponds to the energy of two non-interacting $1s$-excitons,
which is spin independent. Resonances at negative energies $\omega$ correspond
to {\it bound} excitonic molecules (biexcitons). In this case, the
binding energy of the
biexciton is approximately $1.5$ meV. In Fig.~\ref{fig1}~(a), the
masses of electrons and hole are identical (positronium-limit). 
The bound-state of the para-biexciton is the most significant feature at low
energies, since the binding energy is expected to be much smaller 
than the usual
excitonic binding energies, which is $10$ meV in this model for the $1s$-state.
The parallel-spin case has more pronounced spectral weight at lower positive 
energies, but a bound-state is not expected to exist. The spectra have
a maxima at higher energies before dropping to zero.  The bandwidth of
a single-electron band is $50 $ meV. It is surprising that the spectrum
is almost zero above the free electron-hole pair-state bandwidth of
$100$ meV. However, a high energy-resonance is visible.

This distant resonance moves to lower energies in Fig.~\ref{fig1}~(b), 
when the ratio of the electron-to-hole mass is reduced to $m_e/m_h=0.15$, found
in semiconductors like in $GaAs$. The reduced mass of electrons and holes
is kept constant. More spectral features appear in the lower energy
regime. 
In principle, more bound-states should appear for negative energies.
A dip-like structure can be found at almost the same position
for the opposite-spin and parallel-spin case.
In the case of opposite-spin correlations, we find increased spectral
weight for smaller positive energies. 
This can be easily understood, if one eh-pair is in the $1s$-exciton
state, whereas the second eh-pair is quasi-free (dissociated).

This can be seen in Fig.~\ref{fig1}~(c) for infinite hole mass (molecular
limit). The opposite-spin case has pronounced spectral features 
at low energies. More resonances appear, if the spectral broadening in
decreased. The resonances can be classified as
(1) bound excitonic molecules at negative energies, which
simply are the ground-states of two electrons in a static potential of
two {\it heavy} holes with varying distance, and (2) scattering resonances, 
which are anti-binding states with molecular character.
These anti-binding states do contribute to the nonlinear optical
response.  For example, the first large resonance at positive energies
results from the configuration where 
the two holes are extremely close together. This is similar
to the configuration of
a helium atom  from the view of the fast moving electrons.
The spectral weight of the force-force correlation function favors
states with small distance of the charged carriers, as can be seen
from the real-space representation of the state $D^\dagger | 0\rangle$
in Eq.~(\ref{corr}). A third resonance is clearly visible at the same
energy position for both spin cases at $\omega \approx 10$ meV. This
resonance is the largest feature in the case of parallel-spin excitons.
The analysis of small systems of up to $10$-sites strongly suggests that
this feature originates from Coulomb correlation of electrons and
holes in the anti-bonding state, where the holes are located
on neighboring sites. This explains the feature in both spectra.
The position of the resonance is only weakly dependent on the on-site
Coulomb interaction.
We have verified that the spectral resonance from a state with
a single $1s$-exciton and
a dissociated eh-pair gives much a smaller contribution at this energy.
The dip for parallel-spin excitons in Fig.~\ref{fig1}~(c) goes almost
to zero and separates a small band of biexciton scattering states 
with weakly
repulsive interacting pairs of $1s$-excitons from the anti-bonding 
resonance. For opposite-spin excitons, a spectral hole emerges at
zero energy. Spectral weight from the product state of two
$1s$-excitons
with zero center-of-mass momentum is recovered in the mean-field
contribution to the nonlinear optical response.

The force-force correlation spectrum has a large number of spectral
features at the exact biexciton energies, which include 
bound-state as well as scattering-state contributions. The correlation 
function approach treats these states on an equal footing.
%
%
We expect a similar behavior also for more realistic $2d$- or $3d$-models.




%
\section{Application to Four-Wave-Mixing}
\label{section4}
%

In this section, we investigate the correlation effects 
in four-wave-mixing (FWM) experiments with semiconductor heterostructures
in the low-density excitation regime where the third-order theory in the
exciting field is valid. Polarization mixing of electrons and holes
with different spin can be induced and probed using cross-polarized laser
excitation.  With the proliferation of new experiments which 
aim at probing correlations, we use, as illustrations of application of our
theory, two specific experimental situations with
the simple type-I excitation. The first paradigm experiment \cite{exp},
by resonantly exciting excitons in quantum-well structures,
clearly demonstrates the signature of polarization-mixing in these systems
for the ultrafast nonlinear response, even in the absence
of biexcitonic molecules.  The actual numerical simulations are
performed with the quasi one-dimensional semiconductor correlation
function (Sec.~3) to model the spin-dependent effects and are not intended to
quantitatively describe the experiment. As a second example, we 
study the effects of bound and unbound two-exciton states and discuss the
``beating'' phenomena of the biexciton resonance
for type-I  excitations, which was identified as 
quantum-beats \cite{may2} between bound and unbound biexciton states and which
our calculation shows to be a {\it ringing} of the bound-state resonance alone.

%
%

The typical experimental setup is sketched in Fig.~\ref{fig5} for a 
three-pulse four-wave-mixing geometry.
This experiment leads to polarization mixing when pulses (2)
and (3) have opposite helicity. In this case,  equal
populations of excitons with opposite angular momenta are excited for
type-I excitation. In the mean-field description, no third-order polarization
$P^{(3)}({\bf k}_f)$  exists to diffract probe pulse (1) in the
${\bf k}_f = {\bf k}_3 + {\bf k}_2 - {\bf k}_1 $ direction 
for any time delay $T$. However, beyond the mean-field theory, correlation
effects between opposite-spin excitons cause polarization mixing.

To allow for an analytical discussion, we consider the limit of ultrashort 
light pulses with identical central frequency, which 
simplifies most of the following calculations. In this case, the exciting laser
field is given by
\begin{eqnarray}
  {\bf E}(t) = \sum_j   {\bf E}({\bf k}_j) 
\delta(t+\tau_j)   \label{laser}.
\end{eqnarray}
We emphasize that a delta-pulse approximation is used only in the time
integration and not in the frequency integration since it gives a infinite 
broad spectral width. 
The laser pulse in the  propagation direction ${\bf k}_j$ interacts with the
sample at time $t=-\tau_j$. The corresponding {\it area} of the light pulse
is given by ${\bf E}({\bf k}_j)$. The summation index $j$ labels the
different pulses involved the multi-wave experiment.
The first-order polarization of a single exciton transition
is easily calculated from Eq.~(\ref{linpola})
\begin{eqnarray}
  X_{n,\sigma}^{(1)}(t) = i \mu_\sigma \alpha_{n,\sigma}^\ast \sum_j 
  e^{-i(\omega_{n,\sigma}-i\Gamma_\sigma)(t+\tau_j)} \Theta(t+\tau_j)
  E_\sigma({\bf k}_j)   \label{linp}.
\end{eqnarray}
Here, $E_\sigma({\bf k}_j)={\bf E}({\bf k}_j) \cdot {\bf e}_\sigma$
is the projection of the laser field in the propagation direction 
${\bf k}_j$ onto the polarization unit-vector ${\bf e}_\sigma$ 
of the $(n,\sigma)$ transition. The dephasing $\Gamma_\sigma$ may depend on the
polarization of the transition.  In the exciton picture, the total
nonlinear polarization to third-order in the external fields is given
by Eq.~(\ref{pola_result}). The contribution from the phase-space-filling, 
well documented in the literature \cite{yaji,exho,sr}, 
 plays only a minor role in the low-density limit.
We focus on the remaining contribution from the mean-field
part and the genuine correlation which can be treated on equal
footing. Correlation lead to the following nonlinear
complex polarization for the transition $(n,\sigma)$
\begin{eqnarray}
  P^{(3)}_{n,\sigma} (t)  &=& -\frac{i}{2}  
  \mu_\sigma^\ast \alpha_{n,\sigma}
  e^{-i(\omega_{n,\sigma} - i\Gamma_\sigma)t} 
  \sum_{\tilde{n},
    \tilde{\sigma}, n', \sigma', n'',\sigma''
    \atop{j_1,j_2,j_3}} 
  ({\alpha \cal{E} \phi})      \label{fwm}    \\  
  & \left\{ \right. &
    \Theta(t+\tau_{j_1})  \Theta_{21} \Theta_{32} C_{12}(t) +
    \Theta(t+\tau_{j_2}) \nonumber \Theta_{12} \Theta_{32} C_{22}(t) \\ 
                                &+& \left. 
    \Theta(t+\tau_{j_1}) \Theta_{31} \Theta_{23} C_{13}(t)  +
    \Theta(t+\tau_{j_3}) \Theta_{13} \Theta_{32} C_{33}(t) \right\} .\nonumber
\end{eqnarray}

with $\Theta_{kl}=\Theta(\tau_{k}-\tau_{l})$ and an exciton wave-function
dependent factor
\begin{eqnarray}
  \alpha \equiv 
  \alpha_{\tilde{n},\tilde{\sigma}} 
\alpha^\ast_{n',\sigma'} 
  \alpha^\ast_{n'',\sigma''},
\end{eqnarray}
with $\alpha_{n,\sigma}$ defined in Eq.~(\ref{alpha}).
 The exciton-label dependence of $\alpha$ on the left is understood. 
The external field and helicity dependence is contained in the factor
\begin{eqnarray}
  {\cal E} \equiv  \mu_{\tilde{\sigma}} 
  E^\ast_{\tilde{\sigma}}({\bf k}_{j_1})
  \mu_{\sigma'} E_{\sigma'}({\bf k}_{j_2})    \label{field}
  \mu_{\sigma''}E_{\sigma''}({\bf k}_{j_3}).
\end{eqnarray}
A general phase $\phi$ is due to the delay between the short pulses
\begin{eqnarray}
  \phi \equiv
  e^{i(\omega_{\tilde{n},\tilde{\sigma}} 
    + i\Gamma_{\tilde{\sigma}})\tau_{j_1}}
  e^{-i(\omega_{n',\sigma'} - i\Gamma_{\sigma'})\tau_{j_2}}
  e^{-i(\omega_{n'',\sigma''} - i\Gamma_{\sigma''})\tau_{j_3}} \;.
\end{eqnarray}
The nontrivial part of the polarization dynamics in contained
in the time-dependent function $(t>-\tau_a, \tau_a < \tau_b)$
\begin{eqnarray}
  C_{a,b}(t) = \int_{-\tau_a}^t 
  e^{i(\omega_{n,\sigma} 
    + \omega_{\tilde{n},\tilde{\sigma}}  - \omega_{n',\sigma'} 
    -\omega_{n'',\sigma''})t'}  
  e^{-(\Gamma_{\sigma'} + \Gamma_{\sigma''})t'}  
  \int_{0}^{t'+\tau_b} \tilde{F}(\tau) d\tau dt'
  \label{cfun3}.
\end{eqnarray}
Eq.~(\ref{cfun3}) contains the complete Coulomb correlation beyond the
mean-field approximation. The integral kernel is the memory function, which can
be calculated  with the knowledge of the exciton-exciton correlation function,
\begin{eqnarray}
  F_{\tilde{n},\tilde{\sigma};n,\sigma}^{n',\sigma';n^{''},\sigma^{''}}
  (\tau) &=& 
  e^{-i(\omega_{n',\sigma'} + \omega_{n'',\sigma''})\tau} 
  \tilde{F}(\tau).   \label{tilly}
\end{eqnarray}
For the numerical calculation, we always use the resonant condition
for the temporal evolution of $\tilde{F}(\tau)$
in Eq.~(\ref{tilly}), i.e. $\omega_{n'} + \omega_{n''}=0$.
For non-resonant excitation, the
detuning-dependent integrand in Eq.~(\ref{cfun3})  has no phase dependence for
the {\it diagonal} contributions of the correlation function.

We now consider the response from the non-degenerate
exciton ground-states (1s) only, with near-resonant excitation of the central
laser frequency, i.e. $\omega_{n=1s,\sigma} = \omega_{\sigma}$,
 and neglect the transition label $n=1s$. 
We also set $\Gamma_\sigma=\Gamma$ for convenience. This 
simplifies the general expression Eq.~(\ref{fwm}) and
Eq.~(\ref{cfun3}) considerably:
\begin{eqnarray}
  C_{a,b}(t) = \int_{-\tau_a}^t 
  e^{-2 \Gamma t'}  
  \int_{0}^{t'+\tau_b} \tilde{F}(\tau) d\tau
  \hspace{2.cm} (t>-\tau_a, \tau_a < \tau_b)   \label{cfun2}.
\end{eqnarray}

The simplest experiment which is able to distinguish equal-spin from
opposite-spin correlation is the three-beam experiment of
Fig.~\ref{fig5}, where 
the two pulses (2) and (3) interact with the sample at the same time. 
Assuming that lasers (2) and (3) interact with the
sample at $\tau_2 = \tau_3 = 0$ we define the delay time $\tau_1 = T$.
The correlation function is now diagonal with respect to the
exciton indices and has the spectral representation 
with $f_m = \langle 0|D |E_m \rangle $
\begin{eqnarray}
  \tilde{F}(\tau) = \sum_m |f_m|^2 e^{-i\omega_m \tau}
\end{eqnarray}
where the $m$-summation includes all contributions from bound $(\omega_m<0)$
and unbound $(\omega_m>0)$ biexciton states $ |E_m\rangle $ and the
implicit transition indices are understood.
We find for Eq.~(\ref{cfun2}) with $\tau_a=T$, $\tau_b=0$
for negative time delay $T<0$ and $t>-T$,
\begin{eqnarray}
  \tilde{C}(t,T)  &:=&  \sum_m \int_{-T}^t e^{-2\Gamma t'}\int_{0}^{t'}
  |f_m|^2 e^{-i\omega_m \tau} \nonumber \\  
 &=& \sum_m |f_m|^2 \left\{ 
    \frac{ e^{-(2\Gamma +i\omega_m)t} - e^{(2\Gamma +i\omega_m)T} 
      }{i\omega_m(i\omega_m + 2\Gamma)} -
    \frac{ e^{-2\Gamma t} - e^{2\Gamma T}} 
    {i\omega_m 2\Gamma} \right\}  \label{final}.
\end{eqnarray}
                                %
                                %
                                %
%
%
%
The second term of Eq.~(\ref{final}) is simplified by a sum rule
\begin{eqnarray}
  \sum_m |f_m|^2 \omega_m^{-1} 
  =  \beta  \label{sum},
\end{eqnarray}
which can easily be derived from the usual Lehmann representation of
the correlation function and using Eq.(\ref{mfpara}).
%
%
The parameter $\beta$ is the mean-field exciton-exciton interaction parameter
with four identical exciton indices (1s). 
For the parallel-polarized case, this contribution is canceled exactly by the
explicit mean-field contribution to the third-order nonlinear polarization. 
This can be seen by inspection of the second and third
terms on the rhs of Eq.~(\ref{pola_result}), when we set $F(\tau)
\equiv i\beta \delta(\tau)$. For opposite-spin excitation,
this parameter $\beta$ is zero.
In the following we define $C(t,T)$ to be
the result of Eq.~(\ref{final}) after the cancellation:
\begin{eqnarray}
  C(t,T) &:=& \sum_m |f_m|^2  e^{(2\Gamma + i\omega_m)T} 
  \left\{ \frac{ e^{-(2\Gamma +i\omega_m)(t+T)} - 1
    }{i\omega_m(i\omega_m + 2\Gamma)} \right\}
  \label{cfun}
\end{eqnarray}
and, for comparison, the mean-field part
\begin{eqnarray}
  C_{MF}(t,T) &:=& -i \beta e^{2\Gamma T}
  \left\{ \frac{ e^{-2\Gamma (t+T)} - 1}{2\Gamma} \right\}
  \label{cfunmf}.
\end{eqnarray}

The sum-rule Eq.~(\ref{sum}) does not imply that the mean-field
instantaneous contribution in the nonlinear response has completely disappeared
in Eq.~(\ref{cfun}). Mean-field and correlation contributions are treated here 
on an equal footing as part of the Coulomb interaction of the charged carriers.
In the response function Eq.~(\ref{sum}), the mean-field contribution is
recovered in the {\it large} dephasing limit, i.e., $\Gamma>>\omega_x$.
In the following, we assume equal field strength for all pulses 
with real amplitude $E_\sigma$.
We evaluate the signal according to the following general rules:
\begin{enumerate}
\item {\it Spatial dependence}.

  The external sources, Eq.~(\ref{field}), for the nonlinear polarization
  Eq.~(\ref{fwm})
  have to be selected with the correct spatial phase dependence.
  For the signal in ${\bf k}_f$-direction, the index combinations
  ${\bf k}_{j_2}={\bf k}_2, {\bf k}_{j_3}={\bf k}_3$ and
  ${\bf k}_{j_2}={\bf k}_3, {\bf k}_{j_3}={\bf k}_2$ are possible
  with ${\bf k}_{j_1}={\bf k}_1$ fixed. 
\item {\it Time dependence}.

  The set $\{j_1,j_2,j_3 \}$ determines the type of
  response term $C_{ij}$ and the corresponding $\Theta$-functions.
  This will determine the temporal details of the signal, depending
  on the time-order of the incoming pulses.
\item {\it Helicity dependence}.

  The polarization of the transitions and the helicity of the 
  exciting fields
  determine
  the correct type of correlation function and field amplitude
  projection $E_\sigma$.
\item {\it Transition dependence}.

  Perform the summation over the exciton quantum numbers
  after the polarization dependence is determined in the above steps.
  Symmetry arguments can be used to reduce the actual number of 
  terms.
\end{enumerate}

Applying rules $(1)$ and $(2)$ in the case of a near-resonant excitation of the
heavy-hole/light-hole
$1s$-excitons, we find for the time-resolved nonlinear
polarization, using $\mu_\sigma \equiv \alpha_{n=1s,\sigma}^\ast \mu_\sigma$
\begin{eqnarray}
  P^{(3)}_\sigma(t)  &=& - \frac{i}{2}
  \mu_\sigma^\ast 
  \sum_{\sigma', \tilde{\sigma}, \sigma''}
  e^{-i(\omega_\sigma - i\Gamma)t}   
  e^{i(\omega_{\tilde{\sigma}} + i\Gamma)T}
  \mu_{\tilde{\sigma}}^\ast E^\ast_{\tilde{\sigma}}({\bf k}_1)
  \label{tr-pola}
  \\ &\times&
  \left( \rule[-2mm]{0cm}{0.0cm}
    \mu_{\sigma'} E_{\sigma'}({\bf k}_{2})  
    \mu_{\sigma''} E_{\sigma''}({\bf k}_{3}) 
    +
    \mu_{\sigma'} E_{\sigma'}({\bf k}_{3})  
    \mu_{\sigma''}E_{\sigma''}({\bf k}_{2}) 
  \right)  \kappa(t,T)   \nonumber
\end{eqnarray}
with
\begin{eqnarray}
  \kappa(t,T) =
  \Theta(t)  \Theta(T) C(t,0) +
  \Theta(t+T)   \label{kappa}
  \Theta(-T) C(t,T).
\end{eqnarray}
The helicity of the diffracted polarization depends
on the individual contributions, as {\it mixed} in Eq.~(\ref{cfun}).

We discuss in more detail a type-I excitation. 
Corrections due to phase-space filling are neglected. 
The three-pulse experiment in Fig.~\ref{fig5} can clearly distinguish between
correlations between opposite-spin excitons and parallel-spin 
excitons.  In the cross-polarized configuration,
pulses (2) and (3) have opposite  helicity. The response in
${\bf k}_f$-direction has opposite helicity with respect to pulse
(1). This is a consequence of angular-momentum conservation. In the case of
equal helicity, all three pulses and the response have identical
helicity.  In both cases, the main difference comes from the type of heavy-hole
correlation function, which enters the calculation of $C(t,T)$.
Since the response comes from identical optical transitions, besides
the helicity, we set $\mu_\sigma E_\sigma \in \{0,1\}$.
The responses in both polarization configurations differ mainly
in the type of correlation function to be calculated. For the
$\pm$-heavy-hole exciton transition, we find for ${\bf k}_1\rightarrow
\bar{\sigma}$,  ${\bf k}_2 \rightarrow \sigma$ 
and ${\bf k}_3 \rightarrow \bar{\sigma}$,  in the cross-polarized configuration
and ${\bf k}_i \rightarrow \sigma$ in the co-polarized configuration 
$(\bar{\sigma} = -\sigma)$ 
\begin{eqnarray}
  P^{(3)}_{\sigma}(t)  &=& - i 
  e^{-i(\omega_\sigma - i\Gamma)(t+T)}   
  \kappa(t,T) \label{typ1} 
  \left\{ \rule[-2mm]{0cm}{1.0cm} \right. 
  \parbox[h]{6cm}{
    $e^{i(\omega_{\bar{\sigma}} + \omega_{\sigma})T}$ 
    \hspace{1.cm} cross-circular, \\
    $e^{i(\omega_{\sigma} + \omega_\sigma)T}$ \hspace{1.cm} co-circular. } 
\end{eqnarray}
For resonant excitation, i.e., $\omega_\sigma =0$, the time-resolved 
phase of the polarization 
can be read off $P^{(3)} = |P^{(3)}|e^{i\Phi}$ \cite{icps},
\begin{eqnarray}
  \Phi(t) &=& - \Theta(t+T) \Theta(-T) tan^{-1} 
  \left( \frac{Re C(t,T)}{Im C(t,T)} \right)  \\
  &-&
  \Theta(t) \Theta(T) tan^{-1} 
  \left( \frac{Re C(t,0)}{Im C(t,0)} \right) \nonumber,
\end{eqnarray}
in the case where probe pulse (1) comes after the excitation with
pulses (2) and (3),  $T<0$, and 
the case where pulse (1) precedes the excitation, i.e., $T>0$. 
For small times $t$ after the excitation, the instantaneous
phase-space-filling term leads to a $\pi/2$-phase shift of the
polarization with respect to the external field $E$, as discussed
in \cite{chem}.

The observed quantities are the time-resolved (TR) intensity
\begin{eqnarray}
  I^{(3)}(t,T) &=& |P^{(3)}_{\sigma}(t)|^2  \nonumber \\
  &=&
  e^{-2 \Gamma (T+t)} |\kappa(t,T)|^2 \label{tr} 
\end{eqnarray}
and the time-integrated (TI) intensity
\begin{eqnarray}
  I^{(3)}(T) = \int_{-\infty}^{\infty} 
  |P^{(3)}_{\sigma}(t,T)|^2 dt .
\end{eqnarray}
We have performed numerical simulations using a one-dimensional
extended Hubbard model with long-range 
Coulomb interaction as defined in Appendix \ref{apenb}.
We present in this section the exact numerical calculations for this simple
model and  consider in the next section various approximations involving
truncation of the summation in Eq.~(\ref{cfun}) for more complicated models.
For resonant excitation, the ultrafast polarization dynamics is
strongly affected by the relation between the exciton dephasing
parameter $\Gamma$
and the Rydberg energy $\omega_x$, since the detuning is zero.
The Rydberg energy does  not appear explicitly, but for the Coulomb
interaction $\tilde{U}\sim \omega_x$ holds.
The results can be compared
with few-level models on FWM \cite{fwm}.

%
%

The source term $C(t,T)$ of Eq.~(\ref{cfun}) plays a 
central role in the nonlinear response, since it determines the
nontrivial polarization dynamics.
 Many-particle correlation leads to a dynamical
structure which is absent in a simplified non-interacting two-level system. 
 Fig.~\ref{fig7} shows the typical source term for the parallel-spin case.
The mean-field source term shows a finite rise time which
corresponds to the finite rise time of the time-resolved nonlinear polarization
signal, roughly the dephasing time $T_2$. For larger times, the nearly constant
source term leads to an exponential decay of the resulting TR-signal in
Eq.~(\ref{tr-pola}). The mean-field picture is considerably changed when the
exact correlations are taken into account.
Fig.~\ref{fig7} shows the following characteristic features:
(1) an increase in the rise time of the
signal compared to the mean-field approximation,
(2) the signal exhibits a phase dynamics, and (3) the asymptotic 
value is complex and differs considerably from the mean-field value.
Only in the extremely large dephasing limit, not shown in the figure, 
the correlation result approaches the mean-field value $i\beta/2\Gamma$.

%
%

Fig.~\ref{fig6} shows the typical source term for the case of opposite-spin
correlation. The existence of a bound-state biexciton has a strong influence on
the nonlinear response. Oscillations with the biexciton binding frequency of the
single bound-state in the one-dimensional model are visible in Fig.~\ref{fig6}.
The energy denominator in Eq.~(\ref{cfun})
favors low-energy resonances, i.e., isolated bound-states ($\omega_m<0$)
and the low-energy scattering-states continuum.
It is important to note that the oscillations decay with twice
the polarization decay-time $T_2 = \Gamma^{-1}$ in the approximation
of Sec.~\ref{section2}. In the TR-signal, the oscillating
contribution to the signal should, therefore, be fairly small on the decaying
part of the signal.  The second, more important, observation is that,
from Eq.~(\ref{cfun}) and the spectrum in Fig.~\ref{fig1},
the biexciton resonance alone is responsible for the oscillations.
This is a {\it ringing} phenomenon, which is {\it different}
from the usual quantum-beat picture which is suggested by the
few-level model \cite{may2,exp}.

%
%

Fig.~\ref{fig8} shows the results for the time-resolved polarization
for zero delay $T=0$ and weak dephasing. The solid line shows the
mean-field calculation, which obviously gives a poor approximation
for resonant excitation and small dephasing.  The exact result for
cross-polarized circular excitation shows the ringing of the intensity signal 
with the biexciton binding frequency, which is already present in the source term
in Fig.~\ref{fig6}.
The signal for co-polarized excitation is of the same order of
magnitude but shows no oscillatory behavior. The signal peaks at roughly
$T_2/2$, which is less than estimated previously \cite{exp}.

%
%

In Fig.~\ref{fig9}, a shorter dephasing time of $T_2=0.5$ ps is assumed.  
The mean-field result looks better in comparison with the co-polarized
signal.  The exact signals have a decreased rise time of the maximum, 
which can be explained by
an additional dephasing mechanism due to the superposition of the continuum
of two-exciton states, which leads to a natural {\it intrinsic decay},
similar to the effect of an inhomogeneously broadened system. 
%
%
%


%
%

The time-integrated signal in Fig.~\ref{fig10} shows the effect
of  finite delays between the {\it exciting} pulses (2) and (3) and
the {\it probe} pulse (1). The result of the numerical simulation
is in very good agreement with the experimental results by Wang et
al. \cite{exp}, who have performed a three-pulse FWM experiment
on a $GaAs$ quantum-well system. The figure shows the strong decrease
of the co-polarized signal for negative time delay, which is due to the
enhanced intrinsic dephasing of the continuum of two-exciton scattering 
states. The cross-polarized signal is stronger for negative time delay,
which indicates the effect of quasi-bound excitonic molecules.
The spectral weight of the correlation function in Fig.~\ref{fig1}~(b)
for parallel-spin excitons
is more enhanced in the low-frequency regime. This leads to a stronger
total signal in Eq.~(\ref{tr-pola}), 
which can be reproduced for different dephasing parameters.

%
%

Fig.~\ref{fig11} shows the time-integrated intensity of the FWM-signal for 
 co-polarized circular excitation for different dephasing times. For short
dephasing time, we recover the well-known mean-field behavior for
homogeneously broadened systems,  which predicts a rise of the signal $\sim
T_2/4$ and a decay of the signal $\sim T_2/2$ \cite{leo,exho}. This can be
explained by simply counting the number of polarization-waves which are
present before the nonlinear signal is emitted. 
For positive time-delay, even for a longer dephasing time,
the decay $\sim T_2$ can be observed because
correlation effects influence only the strength of the time-integrated
signal. For negative time-delay, significant deviation from
the $\sim T_2/4$-law is found. The probe pulse
interacts with the delay $-T>0$ and fast decaying modes
of the correlations cannot be sustained. 
{\it The calculated results show a smooth transition from a steep rise near zero
delay time, where correlations with the fast modes of the spectrum 
of two-exciton scattering states are important, to a regime where low-energy
modes dominate the response}. The latter again shows the asymptotic
mean-field like $\sim T_2/4$ dependence.

%
%

The most prominent feature of the cross-polarized response in Fig.~\ref{fig12} 
is the {\it modulation} of the signal
with the binding-frequency of the bound-state molecule at negative
time delay. This biexcitonic-effect has been observed experimentally
by various groups \cite{fink,may2} and has sparked
much theoretical effort to improve the mean-field theory of the
semiconductor Bloch equations.  This signature clearly shows 
the importance of correlations, which cannot be neglected for the
resonant excitation of the $1s$-exciton in semiconductor
heterostructures.  No signal in the cross-polarized configuration is predicted
by the mean-field approximation.  For shorter dephasing times, the modulations
disappear very quickly and the signal shows  similar behavior compared to
the co-polarized geometry.
We also observe  a decrease of the modulations, if the pulse-width
is increased. The biexcitonic-modulations (if a bound-state molecule
exists) can only be observed for sufficient short laser pulses and weak
dephasing of the coherence (polarization) in semiconductors. However, none of
the above conditions is necessary to observe correlations due to
the two-exciton scattering-continuum, which is always present and
gives the main contribution, if bound-states are absent because
of impurity-scattering, interface effects, etc.
The correlation function approach gives a unified description of the
observable effects. In principle the same treatment can be applied to type-II
and type-III excitations, but the large number of terms involved
renders the computation quite tedious. Much work still has to be done. A
discussion  of further specific nonlinear optical experiments where 
correlations are involved is in progress.



%

\section{Approximations and comparison with 
related approaches}
\label{section5}

%

In considering possible applications of the correlation function approach for an
improved treatment of the dynamical nonlinear response, 
we need a tractable response theory which takes into account, for example,
(i)  non-resonant, above band-gap excitations where electron-electron
scattering becomes important and where the Markovian approximation is no longer
valid on short time scales,
(ii) coherent and non-coherent scatterings with LO-phonons, 
and (iii) applied magnetic field in heterostructures.
Some theoretical work has already been done in deriving scattering-rate
corrections with memory kernels for the SBE \cite{ban},  in LO-phonon 
corrections \cite{kuz,axt3}, and in high magnetic fields \cite{kner}.
In semiconductor heterostructures or bulk systems, the calculation of the
force-force correlation function, even for resonant excitation, is
an enormous numerical task because of the four-body problem involved. For systems
beyond the simple models for which the exact calculations are possible as
described in the last two sections, we develop various approximation schemes to
incorporate correlation beyond the mean-field level in the dynamical optical
response for low-density excitation. 

\subsection{Excitation induced dephasing (EID)}

EID corrections for the SBE have been discussed by Wang et al. \cite{wang2} 
in an application to FWM, where a phenomenological, density dependent and 
${\bf k}$-diagonal  dephasing-parameter 
was introduced. We can derive a similar correction from the exact
third-order contributions to the nonlinear response.
%
%
In this case, the contribution 
originates entirely from exciton-exciton correlation.

Extracting the fast time-dependence of the linear polarization
in Eq.~(\ref{linpola})
\begin{eqnarray}
  X_{1,n,\sigma}^{(1)}(t) &=&  
  e^{-i \omega_{1,n,\sigma} t}
  \tilde{X}_{1,n,\sigma}^{(1)}(t) 
\end{eqnarray}
and the correlation function, c.f. Eq.~(\ref{tilly}), we find
for the correlation part of Eq.~(\ref{pola_result})
\begin{eqnarray}
&&   
\int_{-\infty}^t 
e^{-2\Gamma (t-t')} 
  F_{\tilde{n},\tilde{\sigma};n,\sigma}^{n',\sigma';
    n^{''},\sigma^{''}}
  (t-t')  
  X^{(1)}_{1,n',\sigma'}(t') X^{(1)}_{1,n^{''},\sigma^{''}}
  (t') dt'        \label{eid}  \\
  =&&    e^{-i (\omega_{1,n',\sigma'} + \omega_{1,n'',\sigma''}) t}
  \int_{-\infty}^t 
  e^{-2\Gamma(t-t')} 
  \tilde{F}_{\tilde{n},\tilde{\sigma};n,\sigma}^{n',\sigma';
    n^{''},\sigma^{''}}
  (t-t')  
  \tilde{X}^{(1)}_{1,n',\sigma'}(t') \tilde{X}^{(1)}_{1,n^{''},\sigma^{''}}
  (t') dt' \nonumber  \\
  \rightarrow& &
  \tilde{\gamma}_{\tilde{n},\tilde{\sigma};n,\sigma}^{n',\sigma';
    n^{''},\sigma^{''}}
  X^{(1)}_{1,n',\sigma'}(t) X^{(1)}_{1,n^{''},\sigma^{''}}(t)
  \nonumber   .
\end{eqnarray}
Hence, we have identified:
\begin{eqnarray}
  \tilde{\gamma}_{\tilde{n},\tilde{\sigma};n,\sigma}^{n',\sigma';
    n^{''},\sigma^{''}}
  &=&   
  \int_{0}^\infty 
  e^{-2\Gamma \tau} 
  \tilde{F}_{\tilde{n},\tilde{\sigma};n,\sigma}^{n',\sigma';
    n^{''},\sigma^{''}}
  (\tau)  d\tau \; ,
\end{eqnarray}
a microscopic expression for the EID for the low-density optical regime.
The imaginary part of $\tilde{\gamma}$
renormalizes the mean-field parameter $\beta$. The real part
of $\tilde{\gamma}$ leads to a {\it dephasing}
of the nonlinear polarization, which can be seen from
Eq.~(\ref{pola_result}). In addition to the phenomenological treatment
\cite{wang2}, polarization-mixing is now also automatically take into account.

\subsection{Short-time memory approximation}

In the limit of large dephasing, a short-time approximation of the memory kernel
in Eq.~(\ref{eid}) is valid by Eq.~(\ref{mfpara2}),
\begin{eqnarray}
  F_{\tilde{n},\tilde{\sigma};n,\sigma}^{n',\sigma';
    n^{''},\sigma^{''}}
  (\tau)  \rightarrow 
  \gamma_{\tilde{n},\tilde{\sigma};n,\sigma}^{n',\sigma';
    n^{''},\sigma^{''}}
 + {\cal O}(\tau)  \; .
\end{eqnarray}
Thus, we  obtain a correlation-modified complex mean-field parameter, which 
exhibits polarization mixing and leads to EID:
\begin{eqnarray}
  \tilde{\gamma}_{\tilde{n},\tilde{\sigma};n,\sigma}^{n',\sigma';
    n^{''},\sigma^{''}}
  \rightarrow
  \frac{\gamma_{\tilde{n},\tilde{\sigma};n,\sigma}^{n',\sigma';
      n^{''},\sigma^{''}}}{2\Gamma}.  \label{eid2}
\end{eqnarray}
We observe that, for diagonal contributions, e.g., $n=n'=a$
and $\tilde{n}=n''=b$, the expression Eq.~(\ref{eid2})
is positive. Hence, the nonlinear polarization for, say $a$, has
an effective dephasing which depends on the density of species $b$
 given by
\begin{eqnarray}
\Gamma_a   \rightarrow \Gamma_a + \frac{\gamma_{a,b}}{\Gamma_a + \Gamma_b} 
X^\ast_b X_b
> 0.
\end{eqnarray}
Explicit results for the expectation value $\gamma_{a,b}$
are given in Appendix \ref{apena}.
                     
\subsection{Non-interacting excitons approximation}

We propose an approximation scheme for
the correlation function $F$, where we replace the time-evolution
for the $D$-operator Eq.~(\ref{corr}) with the full Hamiltonian $H$
of the biexciton subspace ($N_p=2$) with the free time-evolution
of excitons. For more explicit results, we use the $D$-operator
representation of Eq.~(\ref{dexi}):
\begin{eqnarray}
  F_{\tilde{n},\tilde{\sigma};n,\sigma}^{n',\sigma';
    n^{''},\sigma^{''}}
  (\tau)  &\rightarrow&
  \sum_{\beta,\beta',\alpha,\alpha',\atop{{\bf q}\ne 0,{\bf q}'\ne 0}}    
  \tilde{U}_{\bf q} \tilde{U}_{-\bf q'} 
  a_{n,\alpha}({\bf q}) 
  a_{\tilde{n},\alpha'}(-{\bf q})  
  a_{n',\beta}({-\bf q}')^\ast  
  a_{n'',\beta''}({\bf q}')^\ast 
  \nonumber \\ &\times&   
  \langle 0|
  B_{\alpha,\sigma}({\bf q}) 
  B_{\alpha',\tilde{\sigma}}(-{\bf q})  
  B^\dagger_{\beta,\sigma'}(-{\bf q}')
  B_{\beta',\sigma''}^\dagger ({\bf q}')
  |0\rangle  \label{free} \\ &\times&
  e^{-i (\omega_{\beta,\sigma'}({\bf q}') +
    \omega_{\beta',\sigma''}({\bf q}')    
    )\tau} \nonumber.
\end{eqnarray}
This expression can be simplified further. We note that
the expression Eq.~(\ref{free}) is a non-perturbative result which is 
exact to second order in the Coulomb interaction but
properly includes all Coulomb effects to form excitons.
If no bound-state biexcitons 
are present, this expression should give a reasonable approximation beyond
the short-time memory approximation. 
The commutator in Eq.~(\ref{free}) is calculated in Eq.~(\ref{init}).
Since we are interested in the slowly varying contributions to $F(\tau)$, we
restrict the remaining exciton summations in Eq.~(\ref{free}) 
to diagonal contributions, i.e., $\beta=n''$ and $\beta' = n'$. The summation 
cannot be performed analytically for finite $\tau$. According to
Eq.~(\ref{init}),  we can split the correlation function into three
contributions. The {\it bosonic} part gives
\begin{eqnarray}
  F^{(1)} +  F^{(2)} 
  &=&
  \sum_{{\bf q} \ne 0} \;
  \tilde{U}_{\bf q}^2 
  e^{-i (\omega_{n',\sigma'}({\bf q})
    + \omega_{n'',\sigma''}({\bf q}))\tau} \label{per1}
  \\ &\times&
  \left(
    f^{\tilde{\sigma}}_{\tilde{n},n''}({\bf q})  
    f^{\sigma}_{n,n'}(-{\bf q}) 
    f^{\sigma''}_{n'',n''}({\bf q})^\ast 
    f^{\sigma'}_{n',n'}({-\bf q})^\ast  
    \delta_{\sigma,\sigma'} \delta_{\tilde{\sigma},\sigma''} \right.
  \nonumber \\ 
  &+& \left.
    f^{\tilde{\sigma}}_{\tilde{n},n'}({\bf q})  
    f^{\sigma}_{n,n''}(-{\bf q}) 
    f^{\sigma''}_{n'',n''}(-{\bf q})^\ast 
    f^{\sigma'}_{n',n'}({\bf q})^\ast  
    \delta_{\tilde{\sigma},\sigma'} \delta_{\sigma,\sigma''} \right),
  \nonumber.
\end{eqnarray}
with the functions $f$ defined in Eq.(\ref{aneu}).
Due to the composite character of excitons, the third contribution is
\begin{eqnarray}
  F^{(3)}
   &=& -
  \sum_{{\bf q},{\bf k}, {\bf k'}} \tilde{U}_{\bf q}
  \tilde{U}_{\bf k-k'}
  e^{-i (\omega_{n',\sigma'}({\bf q})
    + \omega_{n'',\sigma''}({\bf q}))\tau} \label{per2}
  \\ &\times&
  \left(
    \delta_{\sigma,\sigma''}^{(c)} \delta_{\tilde{\sigma},\sigma'}^{(c)}
    \delta_{\sigma,\sigma'}^{(v)}  \delta_{\tilde{\sigma},\sigma''}^{(v)}
    f^{\sigma''}_{n'',n''}(-{\bf q})^\ast 
    f^{\sigma'}_{n',n'}({\bf q})^\ast   \right. \nonumber  \\
&+& \left.
  f^{\sigma''}_{n'',n''}(-{\bf q}) 
  f^{\sigma'}_{n',n'}({\bf q})  
    \delta_{\sigma,\sigma''}^{(v)} \delta_{\tilde{\sigma},\sigma'}^{(v)}
    \delta_{\sigma,\sigma'}^{(c)}  \delta_{\tilde{\sigma},\sigma''}^{(c)}
  \right) 
\nonumber \\
  & \times & 
  \left(
    \Phi_{\tilde{n},{\bf k'} + {\bf q}, \tilde{\sigma}}^\ast -
    \Phi_{\tilde{n},{\bf k} + {\bf q}, \tilde{\sigma}}^\ast 
    ) (
    \Phi_{n,{\bf k}, \sigma}^\ast -
    \Phi_{n,{\bf k'}, \sigma}^\ast 
  \right) 
  \Phi_{n'',{\bf k'} + \eta_{\sigma''} \kappa_{\sigma''} {\bf q}, \sigma''} 
  \Phi_{n',{\bf k} + \eta_{\sigma'}{\bf q}, \sigma'} \; .
  \nonumber
\end{eqnarray} 
The Kronecker-delta is restricted to the
conduction $(c)$
or valence-band $(v)$, respectively.  
%
%
%
Eq.~(\ref{free}) is a 
non-perturbative result and corresponds to the summation of a 
class of diagrams
in the expansion of the correlation function. 
This can be useful in
calculating
the response from higher dimensional semiconductor models, which
will be addressed in future work.
The low-frequency
range of the spectrum is  approximated by Eq.~(\ref{free}), which
corresponds to 
the long-time behavior of the correlation function and is dominant in the 
nonlinear optical response if no bound-state molecules are present. 
%
%
Higher-frequency modes naturally decay very fast, which affects the
ultra short-time memory.
Eqs.~(\ref{per1},\ref{per2}) can be calculated numerically with
the exciton wave-functions as input.

\subsection{Comparison with collision terms in the Boltzmann approximation}

Finally, we want to approximate the correlation to second-order
in the Coulomb interaction. This will enable us to make a
qualitative comparison of the correlation-part of the nonlinear
optical response with semi-classical Boltzmann-Equation approaches.
This is done here to third-order in the external field.
We expand the {\it polarization} for a transition $\sigma$ 
and Bloch vector ${\bf k}$
in terms of excitons and inspect the source terms which gives the
correlation corrections only. Since the force-force correlation
function is already second-order in $\tilde{U}_{\bf q}$, we
can use the non-interacting linear polarization
Eq.~(\ref{linpola})
\begin{eqnarray}
  X^{(1)}_{1,n,\sigma}(t) 
  &=& i \mu_{\sigma} \sum_{\bf k} 
  \int_{-\infty}^t \langle E_{n,\sigma}|
  e^{-i(H-i\Gamma)(t-t')} |{\bf k}, \sigma \rangle
  E_\sigma(t')
  dt' \\
  &\rightarrow& i \mu_{\sigma} \sum_{\bf k} \Phi_{n,{\bf k},\sigma}^\ast
  \int_{-\infty}^t 
  e^{-i(\epsilon_{{\bf k},\sigma}-i\Gamma)(t-t')}
  E_\sigma(t')
  dt' 
  \nonumber \\
  &=& \sum_{\bf k} \Phi_{n,{\bf k},\sigma}^\ast 
  p^{(1)}_{{\bf k}, \sigma}(t) 
 \nonumber.
\end{eqnarray}
The correlation source term in Eq.~(\ref{pola_result}) to second order in the
Coulomb interaction is
\begin{eqnarray}
  \left. 
    \dot{p}^{(3)}_{{\bf k}, \sigma}(t)\right|_{scatt}  
  &:=& \sum_n \Phi_{n,{\bf k},\sigma}
  \dot{P}_{n,\sigma}(t) \\
  &=& \frac{1}{2} 
  \sum_{n, \tilde{n}, \tilde{\sigma}, n',\sigma', n^{''},\sigma^{''}
    \atop{ {\bf k}_1, {\bf k}_2, {\bf k}_3}}
  \Phi_{n,{\bf k},\sigma}   \Phi_{\tilde{n},{\bf k}_1,\tilde{\sigma}}
  \Phi_{n',{\bf k}_2,\sigma'}^\ast   \Phi_{n'',{\bf k}_3,\sigma''}^\ast
  \left( p^{(1)}_{{\bf k}_1,\tilde{\sigma}}(t) \right)^\ast
  \nonumber \\ &\times&
  \int_{-\infty}^t 
  e^{-2\Gamma(t-t')} 
  F_{\tilde{n},\tilde{\sigma};n,\sigma}^{n',\sigma';
    n^{''},\sigma^{''}}
  (t-t')  
  p^{(1)}_{{\bf k}_2, \sigma'}(t') p^{(1)}_{{\bf k}_3, \sigma^{''}}
  (t') dt' 
  \nonumber.
\end{eqnarray}
The exciton summations can be performed to yield
\begin{eqnarray}   \label{monster}
  &&
  \sum_{n, \tilde{n}, \tilde{\sigma}, n',\sigma', n^{''},\sigma^{''}
    \atop{ {\bf k}_1, {\bf k}_2, {\bf k}_3}}
  \Phi_{n,{\bf k},\sigma}   \Phi_{\tilde{n},{\bf k}_1, \tilde{\sigma}}
  \Phi_{n',{\bf k}_2,\sigma'}^\ast   \Phi_{n'',{\bf k}_3,\sigma''}^\ast
 \\ 
                & \times & 
  F_{\tilde{n},\tilde{\sigma};n,\sigma}^{n',\sigma';n^{''},\sigma^{''}}
  (t-t')
  p^{(1)}_{{\bf k}_1,\tilde{\sigma}}(t)^\ast
  p^{(1)}_{{\bf k}_2, \sigma'}(t') p^{(1)}_{{\bf k}_3, \sigma^{''}}
  \nonumber \\
&=& 
\sum_{{\bf k}_1, {\bf k}_2, {\bf k}_3, {\bf q}, {\bf q'}} 
\tilde{U}_{\bf q} \tilde{U}_{\bf q'}
\nonumber \\
&\times& \langle 0| 
c^\dagger_{{\bf k}_1,\tilde{\sigma}_1}
c_{{\bf k}_1+{\bf q},\tilde{\sigma}_2}
c^\dagger_{{\bf k},\sigma_1}
c_{{\bf k}+{\bf q},\sigma_2} e^{-iH_0(t-t')}
c^\dagger_{{\bf k}_3+{\bf q}',\sigma_2^{''}}
c_{{\bf k}_3,\sigma_1^{''}}
c^\dagger_{{\bf k}_2+{\bf q}',\sigma_2^{'}}
c_{{\bf k}_2,\sigma_1^{'}}
|0 \rangle \nonumber
\\
&  \times &  \left(
  p^{(1)}_{{\bf k}_1,\tilde{\sigma}}(t)^\ast
  p^{(1)}_{{\bf k}_2, \sigma'}(t') 
  p^{(1)}_{{\bf k}_3, \sigma^{''}}
  - p^{(1)}_{{\bf k}_1-{\bf q},\tilde{\sigma}}(t)^\ast
  p^{(1)}_{{\bf k}_2, \sigma'}(t') 
  p^{(1)}_{{\bf k}_3, \sigma^{''}}  \right.  \nonumber \\
&-& p^{(1)}_{{\bf k}_1,\tilde{\sigma}}(t)^\ast
p^{(1)}_{{\bf k}_2-{\bf q}', \sigma'}(t') 
p^{(1)}_{{\bf k}_3, \sigma^{''}}
- p^{(1)}_{{\bf k}_1,\tilde{\sigma}}(t)^\ast
p^{(1)}_{{\bf k}_2, \sigma'}(t') 
p^{(1)}_{{\bf k}_3-{\bf q}', \sigma^{''}}   \nonumber \\
&+& p^{(1)}_{{\bf k}_1,\tilde{\sigma}}(t)^\ast
p^{(1)}_{{\bf k}_2-{\bf q}', \sigma'}(t') 
p^{(1)}_{{\bf k}_3-{\bf q}', \sigma^{''}}
+ p^{(1)}_{{\bf k}_1-{\bf q},\tilde{\sigma}}(t)^\ast
p^{(1)}_{{\bf k}_2, \sigma'}(t') 
p^{(1)}_{{\bf k}_3-{\bf q}', \sigma^{''}}   \nonumber \\
&+& \left. 
p^{(1)}_{{\bf k}_1-{\bf q},\tilde{\sigma}}(t)^\ast
p^{(1)}_{{\bf k}_2-{\bf q}', \sigma'}(t') 
p^{(1)}_{{\bf k}_3, \sigma^{''}}
-
p^{(1)}_{{\bf k}_1-{\bf q},\tilde{\sigma}}(t)^\ast
p^{(1)}_{{\bf k}_2-{\bf q}', \sigma'}(t') 
p^{(1)}_{{\bf k}_3-{\bf q}', \sigma^{''}}  \right) \nonumber \\
&-& ({\bf k} \rightarrow {\bf k}- {\bf q}) \; , \nonumber
\end{eqnarray}
with the non-interacting Hamiltonian $H_0$ being used
for the time evolution.  The matrix element in Eq.~(\ref{monster}) 
is divided into $4$ contributions with a total of $64$
terms, which can be simplified by symmetry arguments.
Note that the {\it bosonic} contributions
to the matrix element yield a $\delta_{{\bf q},\pm{\bf q'}}$
factor. This reproduces the third-order limit of the 
collision terms which were previously derived \cite{ban}. This also corresponds 
to the {\it two-loop} diagrams in the diagrammatic approach \cite{diag1}.
However, the true {\it fermionic} contributions
also appear, which are relevant for the parallel-spin and 
coupled-spin case and correspond to the {\it one-loop} diagrams.
These terms might not be present in the usual semi-classical 
treatment of the scattering rates. 
The correlation function approach naturally incorporates
these effects on the four-particle level and gives
the exact low-density results.

%

\section{Conclusion}
\label{section6}

In this paper, we have presented a {\it unified} theory of
exciton-exciton interaction effects in the third-order nonlinear optical
response, using  a correlation function approach \cite{th1}.
The {\it electronic} problem (dynamics of the four interacting particles) is 
separated from the nonlinear {\it optical} problem.  Furthermore, the correlation
effects beyond the mean-field terms is explicitly represented by a two-exciton
force-force correlation function.  By means of this formalism, we are able to
investigate the role of exciton-exciton  correlations in the third-order
polarization in an application to  resonantly excited heavy-hole excitons in a
semiconductor quantum well. The correlation functions are calculated numerically
for a one-dimensional semiconductor model with long-range Coulomb interaction,
without perturbative approximation.  Their spectra exhibit isolated resonances
due to bound-state biexcitons and continuum  of two-exciton scattering states.
Additional, more pronounced features appear for decreasing mass-ratio of
electron to hole.

A three-pulse FWM configuration can distinguish between parallel-spin and
opposite-spin  correlations. 
For co-polarized excitations, we find a significant deviation from the
expected mean-field $\sim T_2/4$ rise of the time-integrated
FWM-signal for negative time delays between the pulses in the weak dephasing
regime.  This can be explained by an ultrafast intrinsic decay of
correlations due to two-exciton scattering states, since bound-state molecules
are {\it absent} in the case of parallel-spin correlations.
For strong dephasing, we recover the well known mean-field results \cite{exho}. 
The signal in the cross-polarized  configuration is dominated by a modulation
with the binding frequency of the bound-state biexciton in the system, which has
been observed in various experiments \cite{fink,may2}.
These oscillations appear as a ringing of the biexciton mode in the time-resolved
signal, as distinct from true quantum-beats. For positive time delays, the
third-order response is close to the mean-field $T_2/2$-decay behavior.
Polarization-mixing  for cross-polarized excitations is purely a correlation
effect \cite{diag1}, absent in the conventional mean-field approach using the 
semiconductor Bloch equations \cite{2,koch88}. 

We have also derived generalized effective equations of motion for the
polarization for laser excitation near
the fundamental exciton resonance or the exciton continuum states.
This is intended for a future application 
of the theory to more realistic semiconductor quantum-wells.
The exact third-order equations are compared with the results
of Boltzmann-type scattering corrections to the 
quantum-kinetic equations. The effective parameters entering the
dynamics can be calculated in second order of the exciton-exciton
interaction, provided only that the exciton states are known, which
is possible for a large number of semiconductor systems.

The correlation function approach
can be generalized to include additional interactions, 
such as spin-flipping scattering processes or the coupling to
LO-phonons \cite{axt3}.  A further application of the correlation theory 
to spin-beating phenomena in diluted magnetic
semiconductors for the pump-and-probe configuration \cite{awsch}
has been made \cite{th2}.

\acknowledgments

One of the authors (Th. \"O.) acknowledges financial support by the Deutsche
Forschungsgemeinschaft (DFG). This work was supported in part by the
Sonderforschungsbereich SFB $345$, G\"ottingen, and in part by the NSF
Grant No. DMS 94-21966. We also wish to thank
M. Z. Maialle, R. G. Ulbrich, G. B\"ohne and D. S. Chemla for 
useful discussions and comments.

\appendix




%
\section{Commutator algebra for the force operator}
\label{apena}
%

In this Appendix we derive the operators and parameters defined in section
\ref{section2} in terms of a semiconductor Hamiltonian, $H\equiv H_0 + U$, with
an independent-electron part:
\begin{eqnarray}
  H_0 = \sum_{{\bf k}, s} 
  \epsilon_{{\bf k}, s} c_{{\bf k}, s}^\dagger 
  c_{{\bf k}, s} \;,
\end{eqnarray}
where $c_{{\bf k}, s}^\dagger$ creates a Bloch electron
with combined band and spin index $s$ at wave-vector ${\bf k}$, and an
electron-electron interaction term:
\begin{eqnarray}
  U = \frac{1}{2}
\sum_{{\bf k}, s, {\bf k}', s', {\bf q}\ne 0}  
\tilde{U}_{{\bf q},s,s'}
c_{{\bf k}+{\bf q}, s}^\dagger 
c_{{\bf k}'-{\bf q}, s'}^\dagger 
  c_{{\bf k}', s'} 
  c_{{\bf k}, s} \; ,
\end{eqnarray}
where $\tilde{U}_{{\bf q},s,s'}$ is the Coulomb matrix element.

The exciton operator $B_{n,\sigma}$ of a given transition $\sigma$ is associated
with the relative motion wave function $\phi_{n,{\bf k},\sigma}$
at zero center-of-mass momentum:
\begin{eqnarray}
  B_{n,\sigma} =
\sum_{\bf k} \phi_{n,{\bf k},\sigma}^\ast
  c_{{\bf k}, s'}^\dagger 
  c_{{\bf k}, s}. \label{ex0}
\end{eqnarray}
The transition $\sigma=\sigma(s,s')$ connects an electron from a valence band
with combined band and spin label $s'$ to a conduction  band state with label
$s$. The corresponding {\it pair}-operator 
$c_{{\bf k}, s'}^\dagger  c_{{\bf k}, s}$ is denoted in
Sec.~\ref{section2} as $\psi_{{\bf k},\sigma}$.
Optical selection rules determine whether the $\sigma$-transition
is an optically allowed dipole transition with matrix element $\mu_{\sigma}$ or 
a so-called {\it dark} transition. 
%
Dark states are connected to optically active states
via a spin-flip process, which is assumed to be very slow on the
time scales of interest. The selection rules also determine the
corresponding helicity of the dipole transition.

The first step leading to the force operator $D$ is the commutator
$C_{n,\sigma}=[B_{n,\sigma}, H]$, 
\begin{eqnarray}
  C_{n,\sigma} &=& 
  \sum_{{\bf k}} (\epsilon_{{\bf k},s_2} - \epsilon_{{\bf k},s_1})
  \phi_{n,{\bf k},\sigma}^\ast  \label{cdef}
  c_{{\bf k}, s_2}^\dagger c_{{\bf k}, s_1} \\
  &-&
  \sum_{{\bf k}, {\bf q}\ne 0}  
  \tilde{U}_{{\bf q},s_1,s_2} \phi_{n,{\bf k},\sigma}^\ast 
  c_{{\bf k}+{\bf q}, s_2}^\dagger c_{{\bf k}+{\bf q}, s_1} 
  +
  \sum_{{\bf k}, {\bf q} \ne 0}  
  \tilde{U}_{{\bf q},s_1,s_1} \phi_{n,{\bf k},\sigma}^\ast 
  c_{{\bf k}, s_2}^\dagger c_{{\bf k}, s_1} \nonumber
  \\
  &+& \sum_{{\bf k}, {\bf k}', s, {\bf q} \ne 0}  
  \left( \tilde{U}_{{\bf q},s,s_2} \phi_{n,{\bf k}'-{\bf q},\sigma}^\ast - 
    \tilde{U}_{{\bf q},s,s_1} \phi_{n,{\bf k}',\sigma}^\ast 
  \right) 
  c_{{\bf k}+{\bf q}, s}^\dagger c_{{\bf k}, s}
  c_{{\bf k}'-{\bf q}, s_1}^\dagger  c_{{\bf k}', s_2} \; . \nonumber
\end{eqnarray}
Using $c_{{\bf k}, s_1}^\dagger c_{{\bf k}, s_2}=\sum_{\tilde{n}}
\phi_{\tilde{n},{\bf k},\sigma}^\ast 
B_{\tilde{n},\sigma}$, the Wannier equation for the exciton wave function
simplifies the first three terms on the rhs of Eq.~(\ref{cdef}):
\begin{eqnarray}
&  \sum_{\bf k,k'} &  \left[ \left(\epsilon_{{\bf k},s_2} 
- \epsilon_{{\bf k},s_1} 
+ \sum_{{\bf q} \ne 0}
\tilde{U}_{{\bf q},s_1,s_1} \right) \delta_{{\bf k},{\bf k}'}
- \tilde{U}_{{\bf k}-{\bf k}', s_1, s_2} (1-\delta_{{\bf k},{\bf k}'})
\right] 
\Phi_{{\bf k},n,\sigma}^\ast \Phi_{{\bf k}',\tilde{n},\sigma} 
\nonumber \\ 
&=&
\omega_{1,n,\sigma}       
\delta_{n,\tilde{n}} \; ,
\end{eqnarray}
and, therefore,
\begin{eqnarray}
  C_{n,\sigma} &=&  \omega_{1,n,\sigma} B_{n,\sigma} 
\label{cdef2}
  \\
  &+& \sum_{{\bf k}, {\bf k}', s, {\bf q}\ne 0}  
  \left( \tilde{U}_{{\bf q},s,s_2} \phi_{n,{\bf k}'-{\bf q},\sigma}^\ast - 
    \tilde{U}_{{\bf q},s,s_1} \phi_{n,{\bf k}',\sigma}^\ast 
  \right) 
  c_{{\bf k}+{\bf q}, s}^\dagger c_{{\bf k}, s}
  c_{{\bf k}'-{\bf q}, s_1}^\dagger  c_{{\bf k}', s_2} \nonumber .
\end{eqnarray}

The force operator is given by the commutator
$D_{p,l;n,\sigma}=[B_{p,l}, C_{n,\sigma}]$. The first contribution on
the rhs of Eq.~(\ref{cdef2}) vanishes because $[B_{p,l},
B_{n,\sigma}]=0$. 
Thus,
\begin{eqnarray}
  D_{p,l; n,\sigma} &=& 
  \sum_{{\bf k}, {\bf k}', {\bf q}\ne 0}  
  \left( \tilde{U}_{{\bf q},l_2,s_2} \phi_{n,{\bf k}'-{\bf q},\sigma}^\ast 
    - \tilde{U}_{{\bf q},l_2,s_1} \phi_{n,{\bf k}',\sigma}^\ast 
  \right) \phi_{p,{\bf k}+{\bf q},l}^\ast 
    \nonumber \\&-& 
  \left( \tilde{U}_{{\bf q},l_1,s_2} \phi_{n,{\bf k}'-{\bf q},\sigma}^\ast 
-
\tilde{U}_{{\bf q},l_1,s_1} \phi_{n,{\bf k}',\sigma}^\ast 
  \right) \phi_{p,{\bf k},l}^\ast \nonumber \\ &\times&
  c_{{\bf k}+{\bf q}, l_1}^\dagger c_{{\bf k}, l_2}
  c_{{\bf k}'-{\bf q}, s_1}^\dagger  c_{{\bf k}', s_2} \; .
\end{eqnarray}
The spin-independent Coulomb interaction leads to a further simplification.
In terms of the operator
\begin{eqnarray}
  A_{n,\sigma}({\bf q}) &=& 
  \sum_{{\bf k}}
  \left( \phi_{n,{\bf k}-{\bf q}/2,\sigma}^\ast 
    - \phi_{n,{\bf k}+{\bf q}/2,\sigma}^\ast 
  \right)  c_{{\bf k}-{\bf q}/2, s_1}^\dagger c_{{\bf k}+{\bf q}/2, s_2},
\label{aa}
\end{eqnarray}
the $D$-operator can be written in a compact form:
\begin{eqnarray}
  D_{p,l; n,\sigma} &=& 
  \sum_{{\bf q}\ne 0}  
  \tilde{U}_{\bf q} 
  A_{p,l}({\bf q})  
  A_{n,\sigma}(-{\bf q}).    \label{dd}
\end{eqnarray}

The $A$-operator can be related to excitons with finite center-of-mass
momentum ${\bf Q}$. The mass ratio of electrons and holes play a
crucial in the correlation function dynamics. We define
for the transition $\sigma=(s,s')$ the positive mass-ratio 
$\kappa_\sigma=m_{s}/m_{s'}$. With 
\begin{eqnarray}
\eta_{\sigma} = \frac{1}{1+\kappa_{\sigma}}
\end{eqnarray}
the generalization of Eq.~(\ref{ex0}) reads
\begin{eqnarray}
  B_{n,\sigma}({\bf Q}) =
  \sum_{\bf k} \phi_{n,{\bf k},\sigma}^\ast
  c_{{\bf k}-\eta_\sigma {\bf Q}, s'}^\dagger 
  c_{{\bf k}+\kappa_\sigma\eta_\sigma {\bf Q}, s}. \label{ex1}
\end{eqnarray}
We have neglected a possible ${\bf Q}$-dependence of the
relative-motion wave-function, which is valid for parabolic bands. Transforming
the eh representation in Eq.~(\ref{aa}) to one in terms of the excitons leads to
\begin{eqnarray}
  A_{n,\sigma}({\bf q}) &=& 
\sum_{{\bf k},\alpha}
\left( \phi_{n,{\bf k}-\eta_\sigma{\bf q},\sigma}^\ast 
  - \phi_{n,{\bf k}+\eta_\sigma \kappa_\sigma {\bf q},\sigma}^\ast 
\right) \phi_{\alpha,{\bf k},\sigma}  
  B_{\alpha,\sigma}({\bf q}) 
\nonumber \\
&=& 
\sum_{\alpha}
f_{n,\alpha}^{\sigma}({\bf q})
B_{\alpha,\sigma}({\bf q}) \; .
\label{aneu} 
\end{eqnarray}
The $D$-operator in Eq.~(\ref{dd}) becomes in terms of the excitons:
\begin{eqnarray}
  D_{p,l; n,\sigma} &=& 
  \sum_{\alpha,\alpha',{\bf q}\ne 0}  
  \tilde{U}_{\bf q} 
  f_{p,\alpha}^{l}({\bf q})  
  f_{n,\alpha'}^{\sigma}(-{\bf q}) B_{\alpha,l}({\bf q}) 
  B_{\alpha',\sigma}(-{\bf q})  \label{dexi}.
\end{eqnarray}
We use $ \tilde{U}_{\bf q} =  \tilde{U}_{-\bf q} $, to ensure
the symmetry of the $D$-operator in the exciton labels.

The initial value $\tau=0$ of the correlation function
Eq.~(\ref{cfun})
is a ground-state (vacuum)
expectation value. 
We first calculate
\begin{eqnarray}
&&  
\langle 0|
  B_{\alpha',\tilde{\sigma}}({\bf q})  
  B_{\alpha,\sigma}(-{\bf q}) 
  B^\dagger_{\beta,\sigma''}(-{\bf q}')
  B^\dagger_{\beta',\sigma'}({\bf q}')
  |0\rangle    \label{init} \\ &=&
  \delta_{\alpha',\tilde{\sigma};\beta,\sigma''}
  \delta_{\alpha,\sigma;\beta',\sigma'}
  \delta_{{\bf q},-{\bf q}'}
  \nonumber + 
  \delta_{\alpha',\tilde{\sigma};\beta',\sigma'}
  \delta_{\alpha,\sigma;\beta,\sigma''}
  \delta_{{\bf q},{\bf q}'}  
  \\ &+& 
  \langle 0|
  B_{\alpha',\tilde{\sigma}}({\bf q})  
  \left[ [B_{\alpha,\sigma}(-{\bf q}),
  B^\dagger_{\beta,\sigma''}(-{\bf q}')],
  B^\dagger_{\beta',\sigma'}({\bf q}') \right]
  |0\rangle.  \nonumber
\end{eqnarray}
The initial value 
$\gamma_{\tilde{n},\tilde{\sigma};n,\sigma}^{n',\sigma';n^{''},\sigma^{''}}$
of Eq.~(\ref{cfun}) can be split in three
contributions. The first term on the rhs of Eq.~(\ref{init})
gives
\begin{eqnarray}
  \gamma^{(1)}
  &=&
  \delta_{\sigma,\sigma'} \delta_{\tilde{\sigma},\sigma''}
  \sum_{\beta, \beta', {\bf q}} \tilde{U}_{\bf q}^2 
  \label{a1} 
  f^{\tilde{\sigma}}_{\tilde{n},\beta}({\bf q})  
  f^{\sigma}_{n,\beta'}(-{\bf q}) 
  f^{\sigma''}_{n'',\beta}({\bf q})^\ast 
  f^{\sigma'}_{n',\beta'}({-\bf q})^\ast.
\end{eqnarray} 
The summation over the exciton labels $\beta,\beta'$ can be performed,
using the abbreviation
\begin{eqnarray}
g^{\sigma}_{n,m}({\bf q}) &=& \sum_\beta
f^{\sigma}_{n,\beta}({\bf q})
f^{\sigma}_{m,\beta}({\bf q})^\ast \\
&=& \sum_{\bf k}
\left( \phi_{n,{\bf k}-\eta_\sigma{\bf q},\sigma}^\ast 
  - \phi_{n,{\bf k}+\eta_\sigma \kappa_\sigma {\bf q},\sigma}^\ast 
\right)
\left( \phi_{m,{\bf k}-\eta_\sigma{\bf q},\sigma}
  - \phi_{m,{\bf k}+\eta_\sigma \kappa_\sigma {\bf q},\sigma} 
\right)  \; , \nonumber    
\end{eqnarray}
which gives
\begin{eqnarray}
  \gamma^{(1)}
  &=&
  \delta_{\sigma,\sigma'} \delta_{\tilde{\sigma},\sigma''}
  \sum_{{\bf q}} \tilde{U}_{\bf q}^2 
  \label{a11} 
  g^{\tilde{\sigma}}_{\tilde{n},n''}({\bf q})  
  g^{\sigma}_{n,n'}({-\bf q})
\end{eqnarray} 
and for the second term on the rhs of Eq.~(\ref{init})
\begin{eqnarray}
  \gamma^{(2)}
  &=&
  \delta_{\sigma,\sigma''} \delta_{\tilde{\sigma},\sigma'}
  \sum_{{\bf q}} \tilde{U}_{\bf q}^2 
  \label{a21} 
  g^{\tilde{\sigma}}_{\tilde{n},n'}({\bf q})  
  g^{\sigma}_{n,n''}({-\bf q}).
\end{eqnarray} 
The third contribution arises only if the two electron-hole
pairs have at least one band in common, i.e., in the parallel-spin or
coupled-spin case, c.f. Sec.~\ref{section3}.
In the case of excitons being ideal {\it bosons}, only $\gamma^{(1)}$ and
$\gamma^{(2)}$ would contribute. We give the result for the
general case:
\begin{eqnarray}
  \gamma^{(3)}
  &=& -
  \left(
    \delta_{\sigma,\sigma''}^{(c)} \delta_{\tilde{\sigma},\sigma'}^{(c)}
    \delta_{\sigma,\sigma'}^{(v)}  \delta_{\tilde{\sigma},\sigma''}^{(v)}
+
    \delta_{\sigma,\sigma''}^{(v)} \delta_{\tilde{\sigma},\sigma'}^{(v)}
    \delta_{\sigma,\sigma'}^{(c)}  \delta_{\tilde{\sigma},\sigma''}^{(c)}
  \right)
  \sum_{{\bf q},{\bf k}, {\bf k'}} \tilde{U}_{\bf q}
  \tilde{U}_{\bf k-k'}
  \label{a31}  \\
  & \times &(
  \Phi_{\tilde{n},{\bf k'} - {\bf q}/2, \tilde{\sigma}}^\ast -
  \Phi_{\tilde{n},{\bf k'} + {\bf q}/2, \tilde{\sigma}}^\ast 
  ) (
  \Phi_{n'',{\bf k'} - {\bf q}/2, \sigma''} -
  \Phi_{n'',{\bf k} - {\bf q}/2, \sigma''} 
  )\nonumber \\ & \times& (
  \Phi_{n',{\bf k} + {\bf q}/2, \sigma'} -
  \Phi_{n',{\bf k'} + {\bf q}/2, \sigma'} 
)(
  \Phi_{n,{\bf k} + {\bf q}/2, \sigma}^\ast -
  \Phi_{n,{\bf k} - {\bf q}/2, \sigma}^\ast 
  ) \nonumber .
\end{eqnarray} 
The Kronecker-delta symbol with an upper index means, the corresponding
conduction $(c)$
or valence-band $(v)$ must be identical for the transition pair.




%
\section{An interacting semiconductor model}
\label{apenb}
%

We present the details of the real-space extended Hubbard-model
used for the numerical calculations of the correlation-function
and the linear optical properties (excitons).
We start with the general $d$-dimensional lattice-model of an electron-hole
system and briefly review the fundamental aspects of the problem on a lattice.
The kinetic energy part of a two-band lattice model is the
usual hopping term
\begin{eqnarray}
H_0 = \sum_{{\bf m},{\bf m}',s} t^{(s)}_{{\bf m},{\bf m}'}
c^\dagger_{{\bf m},s} c_{{\bf m}',s}
+ h.c.,   \label{kin}
\end{eqnarray}
where $  c^\dagger_{{\bf m}',s}$ creates an electron in a Wannier-state
at the lattice site ${\bf m}$ with a combined band and spin index 
$s$ and $ t^{(s)}_{{\bf m},{\bf m}'}$
is the hopping matrix element between two sites. The hopping is restricted to
nearest neighbor  with parameter $t^{(s)}$ for each band. 
We take $t^{(s)}<0$  $(t^{(s)}>0)$ for an s-type (p-type) conduction (valence)
band, allowing for interband optical transitions.
The potential energy of the system is given by the Coulomb interaction between
the electrons and charge-positive lattice ions with charge $e Z_{core}>0$. 
In terms of the dimensionless charge-density (electrons and ions)
at site ${\bf m}$
\begin{eqnarray}
\rho_{\bf m} = \left(\sum_s
c^\dagger_{{\bf m},s} c_{{\bf m},s}\right) - Z_{core} \; ,
\end{eqnarray}
the Coulomb interaction in the Wannier-states to leading order
is given by the electrostatic monopole-monopole contribution
\begin{eqnarray}
\hat{U} = \frac{1}{2} \sum_{{\bf m},{\bf m}'} U_{{\bf m},{\bf m}'}
\rho_{\bf m} \rho_{{\bf m}'} .    \label{cou}
\end{eqnarray}
where $U_{{\bf m},{\bf m}'}=e^2/|{\bf m}-{\bf m}'|$ for different
sites and the on-site Coulomb interaction $U_{{\bf m},{\bf m}}\equiv U_0$
is an additional parameter in the model. 
 
The ground-state $|0\rangle$ of the lattice model is given by 
the completely filled valence-band states at each site. For the numerical
calculation of the correlation function, a four-band approximation of the
semiconductor is sufficient, i.e., only two holes and two electrons
of one or two species are relevant. The index $s$ is split into two bands and
two spin directions. This leads to different types of correlation functions,
 as discussed in  Sec.~\ref{section3}.  We write the ground state as
\begin{eqnarray}
|0\rangle = \Pi_{\bf m}  c^\dagger_{{\bf m},1,\uparrow}
c^\dagger_{{\bf m},1,\downarrow} |vac\rangle .
\end{eqnarray}
The spatial extension of the relative motion of a {\it single}
exciton in its ground state depends on the ratio of the sum of
the bandwidths $B_s\equiv B_1+B_2=2d(t_1-t_2)$ and the nearest
neighbor Coulomb energy $U_1\equiv e^2/a_L$, where $a_L$ is the
lattice constant. For $B_s/U_1 \gg 1$ the results of the lattice
model are similar to the continuum limit given by  
the usual two-parabolic-band effective-mass model. If one expands the band
dispersion quadratically around the center of the Brillouin zone one finds 
for the Bohr radius of the exciton for $d=3$
\begin{eqnarray}
\label{anull}
a_0=\frac{2(t_1-t_2)}{U_1}a_L
\end{eqnarray}
and the exciton binding energy $\omega_x$ is given by
\begin{eqnarray}
\omega_x= \frac{U_1}{2}\left(\frac{a_L}{a_0}\right).
\end{eqnarray}
In the following, we give all energies in units of the Rydberg energy 
$\omega_x$.
One should keep in mind that this is {\it not} the
exciton binding energy in $d=1$, which apart from a different
dimensionless prefactor depends on the ratio $\eta\equiv U_0/U_1$.
The different effective electron (hole) masses enter via $|t_1/t_2|=|m_2/m_1|$.
The total Hamiltonian operator $H=H_0+\hat{U}$
is conveniently transformed into the electron-hole picture
defining $c_{{\bf m},s} = c_{{\bf m},2,s}$ for electron states
and $h_{{\bf m},s} = c^\dagger_{{\bf m},1,s}$ for the hole states.
Let $|\Phi\rangle$ denote a many-particle eigenstate of $H$ with an 
arbitrary number $N_p$ of the eh-pairs as a  quantum-number.
As the corresponding pair-number operator $N_p$ commutes with
$H$, we can always work with eigenstates of $N_p$. This is the
fundamental assumption for the Hubbard-operator formulation of the 
Hamiltonian in Sec.~\ref{section2}.

We now discuss in detail the results of the numerical calculations for a 
quasi-one-dimensional ring model. The Coulomb interaction Eq.~(\ref{cou})
between two charges depends on the chord-distance \cite{solid}
\begin{eqnarray}
|{\bf m}-{\bf m}'|=\frac{N a_L}{\pi} \sin[{\frac{\pi}{N} |m-m'|}]
\end{eqnarray}
of the sites, which are labeled by the dimensionless
numbers $m$. 
The model interpolates smoothly between the two limits of
Frenkel excitons $a_0/a_L \rightarrow 0$ and
the Wannier limit of large extended objects $a_0/a_L\gg 1$, which
is only limited by the finite total system size N.
The system size for the numerical calculation of the correlation
function is $N=120$ for equal-spin correlation and $N=140$ for
opposite-spin correlation.
The on-site Coulomb interaction is
fixed with $\eta=1.5$. We have also fixed the bandwidth
$B:=|4(t_2-t_1)|$ of the eh-pair continuum for each model, which
depends only on the reduced electron and hole mass, while
the positive mass ratio $m_e/m_h$ is varied. The mass
ratio is an important parameter of the two eh-pair subspace.
We compare the positronium limit $m_e/m_h=1$ with
the semiconductor $GaAs$ case of  $m_e/m_h \approx 0.15$ and
the molecular (hydrogen) limit  $m_e/m_h \ll 1$.
The long-range Coulomb potential leads to the formation of a finite
number of bound exciton-states in the system in contrast to the usual
Hubbard-model with purely on-site repulsion $U_0$.

\subsection{Parallel-spin excitons}
The parallel-spin exciton case is characterized by the
exchange repulsion between same type carriers in the bands, which, 
in an ideal one-dimensional system, plays
the role of dynamical boundary conditions.
The parallel-spin case is relevant when the optical excitation process
is limited to a single circularly polarization of the external field.
A complete set of parallel-spin two-pair states with zero
center of mass momentum is given by \cite{solid}
\begin{eqnarray}
|p,\alpha,\beta \rangle  &=& 
\frac{C_p^{(\alpha,\beta)}}{\sqrt{N}}
 \sum_{m}  \label{set}
c^\dagger_{m+p+\alpha} h^\dagger_{m + p} c^\dagger_{m+\beta}
h^\dagger_{m} |0\rangle   , 
\end{eqnarray}
where 
\begin{eqnarray}
C_p^{(\alpha,\beta)} = \left\{ 
 \begin{array}{cc}  
\sqrt{2}  & \;\;\;\; p=N/2 \;\; and \;\; \alpha=\beta \\ 1 & \;\;\;\; else
\end{array} \right.
\end{eqnarray}
is a normalization constant and a triplet
$(p,\alpha,\beta)$ labels the relative position of
the carriers. In order to avoid double counting of states,
the set of possible triplets is restricted. The (single) 
spin-index is understood.

The real-space representation of 
the product state of
two excitons with quantum numbers $a$ and $b$ with
band indices $(a_2,a_1)$ and $(b_2, b_1)$, which 
enters the calculation of the generating
correlation function, is given by
\begin{eqnarray}
  M_{a,b}^\dagger |0\rangle &=& \frac{1}{N} 
  \sum_{m_1,m_2,m_3,m_4}
  \phi_b(m_1-m_2) \phi_a(m_3-m_4)  \label{pro}  \\ &\times&
  c^\dagger_{m_1,b_1} h^\dagger_{m_2,b_2} c^\dagger_{m_3,a_1}
  h^\dagger_{m_4,a_2} |0\rangle  \nonumber,
\end{eqnarray}
where each exciton has zero center-of-mass momentum.
The real-space exciton wave-function is given by                              
\begin{eqnarray}
  \phi_{a,k} = \frac{1}{\sqrt{N}} 
  \sum_{m} e^{-ikm}
  \phi_a(m) \; ,
\end{eqnarray}                                
and the exciton operator can be expressed as    
\begin{eqnarray}
  B_a^\dagger = \frac{1}{\sqrt{N}} 
  \sum_{m_1,m_2} \phi_a(m_1-m_2) c^\dagger_{m_1,a_1} h^\dagger_{m_2,a_2} .
\end{eqnarray}
For the force-force correlation function in
the nonlinear optical response, we need a 
linear superposition of two-pair states in the
initial state $D_{a,b}^\dagger |0\rangle$ 
\begin{eqnarray}
  D_{a,b}^\dagger |0 \rangle &=&  \frac{1}{N} 
  \sum_{m_1,m_2,m_3,m_4}
  \phi_b(m_1-m_2)   \label{corr}  \phi_a(m_3-m_4) 
  \\ &\times&
  \left( U_{m_3-m_1} -  U_{m_4-m_1} -  U_{m_3-m_2} +  U_{m_4-m_2} \right)
  \nonumber  \\ &\times&
  c^\dagger_{m_1,b_1} h^\dagger_{m_2,b_2} c^\dagger_{m_3,a_1}
  h^\dagger_{m_4,a_2} \nonumber   |0\rangle .
\end{eqnarray}
Both states are center-of-mass eigenstates with zero total momentum
in systems with periodic boundary conditions. It is possible to
fix the hole position $m_4$ and to introduce relative distances with
respect to this hole to reduce the number of coordinates. 
The states Eq.~(\ref{pro}) and Eq.~(\ref{corr}) can then be
expressed in the basis set Eq.~(\ref{set}).

\subsection{Opposite-spin excitons}

The corresponding basis set for the opposite-spin pair-states is chosen
as
\begin{eqnarray}
  |m_1,m_2,m_3 \rangle &=&  \frac{1}{\sqrt{N}} 
  \sum_{m_4}
  c^\dagger_{m_3+m_4,a_1} h^\dagger_{m_2 + m_4,a_2} c^\dagger_{m_1 + m_4,b_1}
  h^\dagger_{m_4,b_2} |0\rangle,
\end{eqnarray}
where the opposite-spin indices for the carriers in one band are
understood, i.e., $b_2\ne a_2$ and $b_1 \ne a_1$.
For numerical work, the absence of the Pauli-blocking for the two
electrons/holes in the bands leads to a quite simple counting of opposite-spin
states in contrast to the equal-spin case. 

The numerical evaluation proceeds in the usual way.
The total number of basis states involved in the
calculation is $N^3$ for the opposite-spin problem, 
which gives a vector length of about 
$3\times 10^6$. The Lanczos algorithm requires two states for the
iteration, which reside in memory to speed up the process and
can be handled quite well on a PC with a total of $64$ MByte 
memory. The Lanczos algorithm tridiagonalizes the
Hamiltonian for the $4$-particle problem, starting with the 
(normalized) initial state $D^\dagger 0\rangle$ for the
force-force correlation function. Each iteration step produces one new
basis state. The iteration is extremely fast, due to the sparseness of
the Hamiltonian matrix of $H$ in real-space. We truncate the iteration
after the spectrum of the resolvent matrix $(z-H)^{-1}$ stabilizes.
No eigenvectors or eigenvalues for the biexciton problem
have to be calculated.
The relevant spectrum is given in the usual way
\begin{eqnarray}
  f_{a,b}(\omega) = -2 Im \langle 0 | D_{a,b} \frac{1}{\omega - H + i0} 
  D^\dagger_{a,b}|0   \rangle
\end{eqnarray}
from the inversion of the resolvent matrix, which is simple in
the tridiagonal representation of $H$.

\begin{figure}
    \caption{Selection rules of heavy and light-hole exciton transitions
      in Zincblende structures. A splitting of heavy- and light-hole
      states in heterostructures 
      allows for a selected excitation of these transitions (I).
      The degenerate case (II) is shown for each helicity of the
      exciting field. The $\Lambda$-transition (III) is excited by linear
      polarization.
      \label{fig4}}
\end{figure}

\begin{figure}
  \caption{Force-force correlation function spectra $F(\omega)$ for
    the
    $1s$-exciton contribution for opposite spins (solid line)
    and parallel spins (dashed line).
    The mass-ratio of electrons and holes is $m_e/m_h=1$ for (a), which
    corresponds to the {\it positronium} limit, $m_e/m_h=0.15$ for
    (b), which
    corresponds to the heavy-hole/electron mass ratio of $GaAs$ and
    for the 
    {\it molecular}
    (hydrogen) limit $m_e/m_h=0$ in (c). 
    Bound excitonic molecules appear for $\omega<0$ and 
    continuum two-exciton contribution
    have $\omega>0$. 
    \label{fig1}
    }
\end{figure}

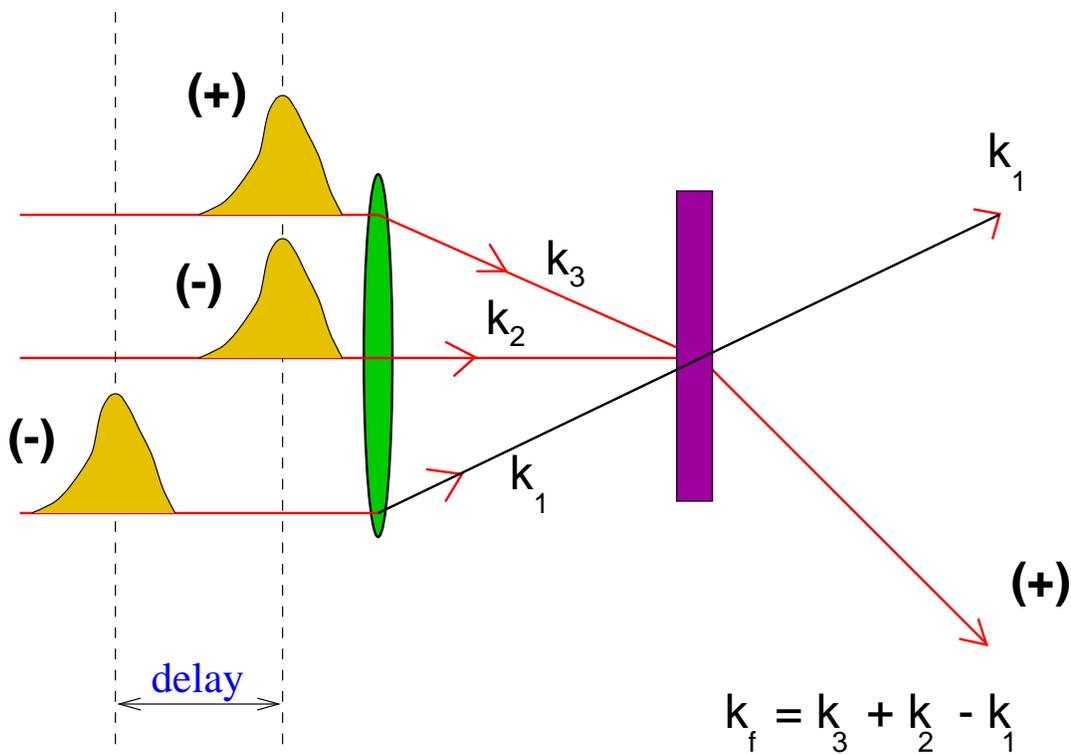
\begin{figure}
  \caption{Three-pulse four-wave-mixing transmission geometry.
    \label{fig5}}
  \end{figure}

\begin{figure}
  \caption{Source term $C(t,0)$ in the nonlinear response
    in the parallel-spin case with parameters of Fig.~\protect\ref{fig6}).
    The mean-field approximation, which has no real part from 
                                Eq.~(\protect\ref{cfunmf}) overestimates the response.
    \label{fig7}
    }  
\end{figure}

\begin{figure}
    \caption{Source term $C(t,0)$ for the nonlinear response
    in the opposite-spin case with the spectral function of 
                                Fig.~\protect\ref{fig1}~(b).
    The dephasing time is $T_2=2$ ps. The pronounced oscillations
    with a period of $2.8$ ps ($E_{xx}=1.5$ meV)
    origin from the bound-state biexciton in the system.
    \label{fig6}
    }
\end{figure}

\begin{figure}
  \caption{TR-signal of the nonlinear polarization for
    the one-dimensional semiconductor model for $T=0$ in the weak
    dephasing
    limit. The oscillations with the biexciton binding energy 
    are a ringing in the signal, since no additional biexciton states
    are necessary for the response. The dephasing time is $T_2=4$~ps.
    \label{fig8}
    }
\end{figure}

\begin{figure}
    \caption{TR-signal of the nonlinear polarization for
    the one-dimensional semiconductor model for $T=0$ in the strong
    dephasing
    limit in comparison with the mean-field response for parallel-spin
    excitation.  The dephasing time is $T_2=0.5$~ps.
    \label{fig9}
    }  
\end{figure}

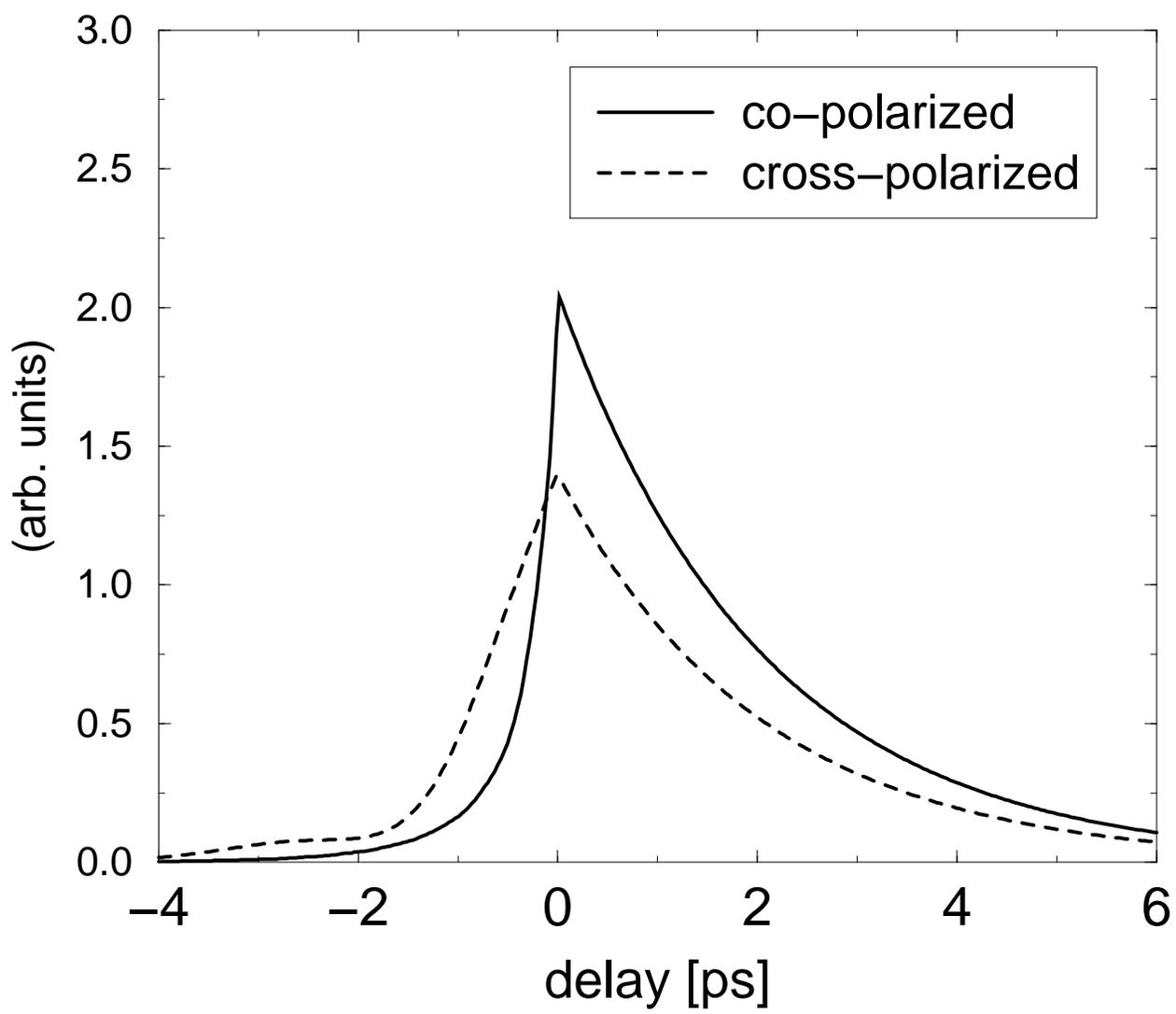
\begin{figure}
  \caption{TI-intensity of the FWM-signal for co-polarized (solid
    line)
    and cross-polarized (dashed line) 
    circular excitation for a dephasing time of $T_2=4$ ps.
    \label{fig10}
    }  
\end{figure}

\begin{figure}
  \caption{Normalized TI-intensity of the FWM-signal for 
    co-polarized  
    circular excitation on a log scale.
    For negative time delay, significant deviation from
    the exponential decay with a rise-time of $T_2/4$ is observed
    for smaller dephasing.
    \label{fig11}
    }  
\end{figure}

\begin{figure}
  \caption{Normalized TI-intensity of the FWM-signal for 
    cross-polarized  
    circular excitation on a log scale.
    For negative  delay times, oscillations with
    the binding energy of the biexciton are visible
    for sufficiently small dephasing.
    \label{fig12}
    }  
\end{figure}

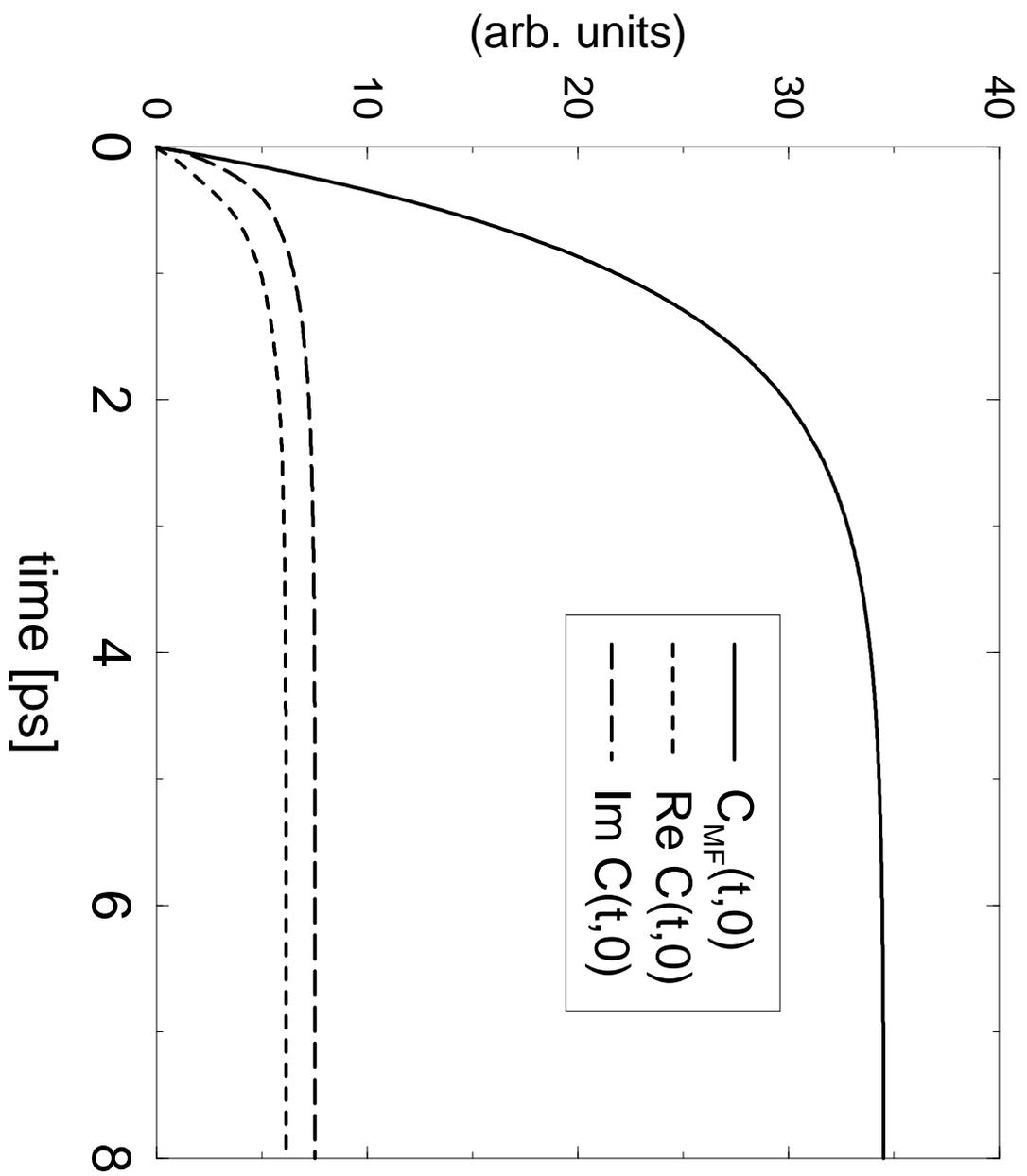
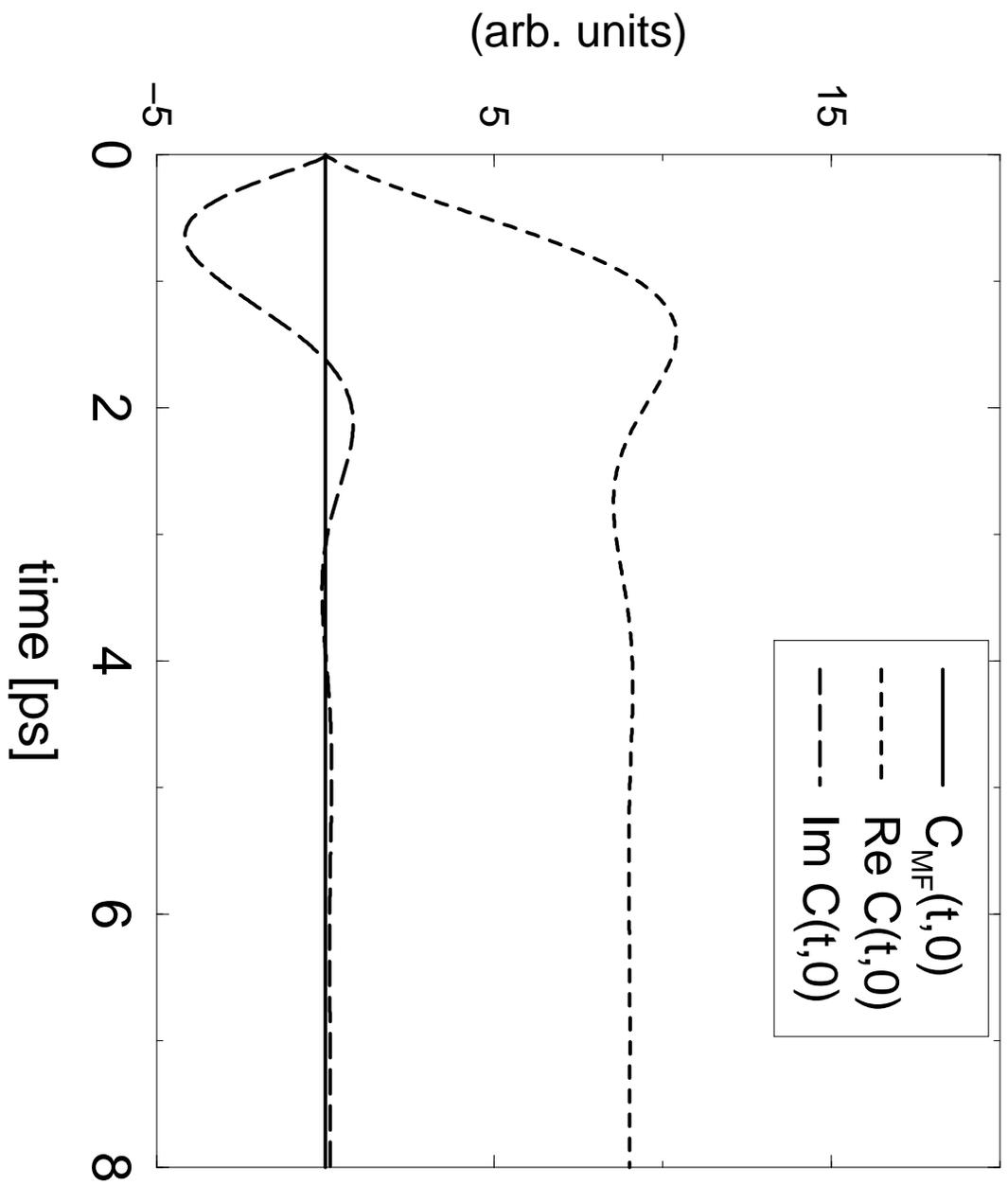
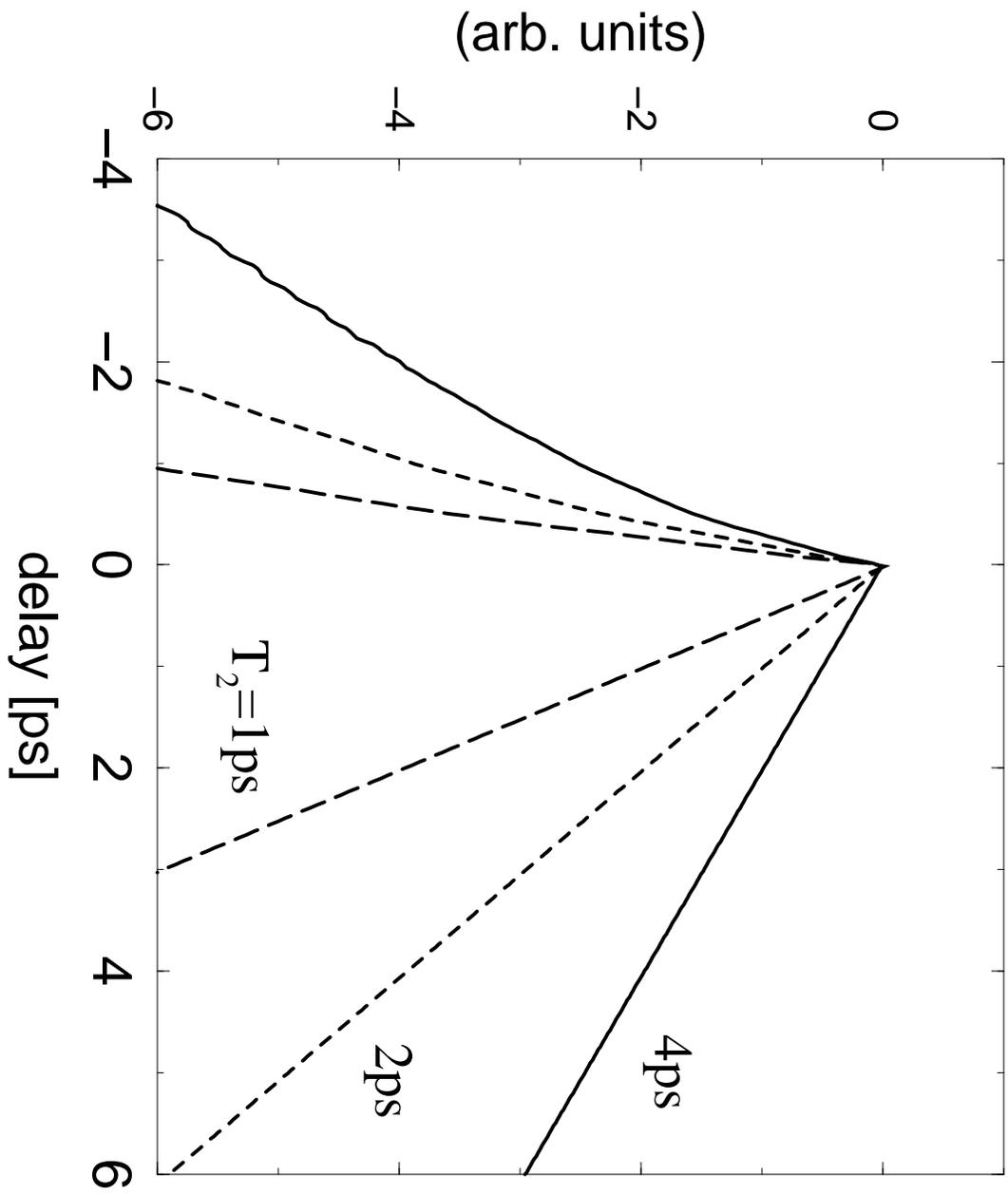
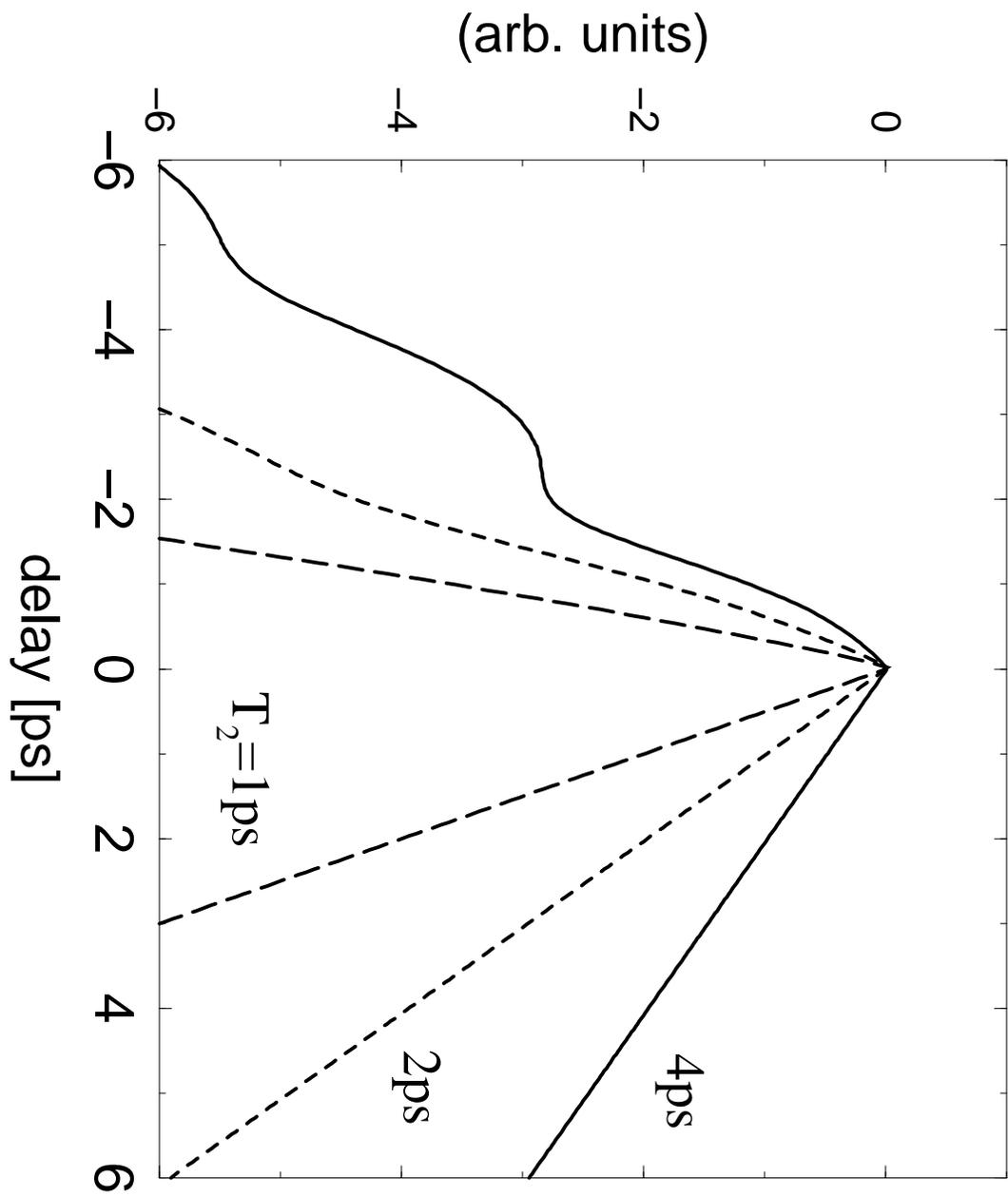
\end{document}